\documentclass[twocolumn]{aastex631}

\usepackage{subfigure}
\usepackage{url}
\usepackage{hyperref}
\usepackage{longtable}
\usepackage{natbib}
\usepackage{amsmath}
\usepackage{listings}
\usepackage[normalem]{ulem}
\usepackage{bm}
\usepackage{comment}
\usepackage{array}
\newcolumntype{P}[1]{>{\centering\arraybackslash}p{#1}}
\newcolumntype{M}[1]{>{\centering\arraybackslash}m{#1}}

\usepackage{xcolor, fontawesome}
\definecolor{twitterblue}{RGB}{64,153,255}

\newcommand{\twitter}[1]{\href{https://twitter.com/#1 }{\textcolor{twitterblue}{\faTwitter}\,\tt \textcolor{twitterblue}{@#1}}}

\newcommand{\github}[1]{\href{https://github.com/#1 }{\textcolor{black}{\faGithub}\,\tt \textcolor{black}{@#1}}}

\definecolor{Code}{rgb}{0,0,0}
\definecolor{Decorators}{rgb}{0.5,0.5,0.5}
\definecolor{Numbers}{rgb}{0.5,0,0}
\definecolor{MatchingBrackets}{rgb}{0.25,0.5,0.5}
\definecolor{Keywords}{rgb}{1,0,0}
\definecolor{self}{rgb}{0,0,0}
\definecolor{Strings}{rgb}{0,0.63,0}
\definecolor{Comments}{rgb}{0,0.63,1}
\definecolor{Backquotes}{rgb}{0,0,0}
\definecolor{Classname}{rgb}{0,0,0}
\definecolor{FunctionName}{rgb}{0,0,0}
\definecolor{Operators}{rgb}{0,0,0}
\definecolor{Background}{rgb}{0.98,0.98,0.98}
\definecolor{Booleans}{rgb}{0.572,0,0.572}
\definecolor{BuiltinFunction}{rgb}{0.572,0,0.572}
\definecolor{BuiltinConstant}{rgb}{0.572,0,0.572}
\definecolor{Asterisk}{rgb}{0.670,0,1}

\lstdefinelanguage{Python}{
 	numbers=left,
 	numberstyle=\footnotesize,
 	numbersep=7pt,
 	xleftmargin=1.26em,
 	framextopmargin=2em,
 	framexbottommargin=2em,
 	showspaces=false,
 	showtabs=false,
 	showstringspaces=false,
 	frame=l,
 	tabsize=4,
 	stepnumber=1,
	basicstyle=\small\ttfamily,
 	backgroundcolor=\color{Background},
	stringstyle=\ttfamily\color{Strings},
	morekeywords={import,from,class,def,while,if,in,elif,else,not,or,print,break,continue,return,access,as,except,exec,finally,global,import,lambda,pass,print,raise,try,assert},
 	keywordstyle={\color{Keywords}\bfseries}, 
	otherkeywords={[2]*},
	keywordstyle={[2]\color{Asterisk}},
}

\usepackage{color}

\newcommand{\shrug}{\texttt{\raisebox{0.75em}{\char`\_}\char`\\\char`\_\kern-0.5ex(\kern-0.25ex\raisebox{0.25ex}{\rotatebox{45}{\raisebox{-.75ex}"\kern-1.5ex\rotatebox{-90})}}\kern-0.5ex)\kern-0.5ex\char`\_/\raisebox{0.75em}{\char`\_}}}

\newcommand{\nflares}{13}

\definecolor{linkcolor}{rgb}{0.41960784, 0.40784314, 0.39607843}
\newcommand{\codeicon}{{\color{linkcolor}\faCloudDownload}}

\newcommand{\angstrom}{\mbox{\normalfont\AA}}
\newcommand{\lya}{Ly-$\alpha$}

\newcommand{\rockcliffep}{Rockcliffe et al. in prep}

\newcommand{\calcos}{\texttt{calcos}}
\newcommand{\costools}{\texttt{costools}}

\newcommand{\aumic}{AU~Mic}
\newcommand{\planetb}{AU~Mic\,b}
\newcommand{\planetc}{AU~Mic\,c}
\newcommand{\planetboth}{AU~Mic\,b~and~c}

\newcommand{\chicago}{Department of Astronomy and Astrophysics, University of
Chicago, 5640 S. Ellis Ave, Chicago, IL 60637, USA}
\newcommand{\lab}{Laboratory for Atmospheric and Space Physics, University of Colorado, 600 UCB, Boulder, CO 80309, USA}

\newcommand{\nsf}{NSF Graduate Research Fellow}

\begin{document}
\title{AU Microscopii in the FUV: Observations in Quiescence, During Flares, and Implications for \planetboth}

\shorttitle{AU Mic FUV Flares} 
\shortauthors{Feinstein et al.}

\author[0000-0002-9464-8101]{Adina~D.~Feinstein}
\altaffiliation{\nsf}
\affiliation{\chicago}

\author[0000-0002-1002-3674]{Kevin~France}
\affiliation{\lab}
\affiliation{Department of Astrophysical and Planetary Sciences, University of Colorado, UCB 389, Boulder, CO 80309, USA}
\affiliation{Center for Astrophysics and Space Astronomy, University of Colorado, 389 UCB, Boulder, CO 80309, USA} 

\author[0000-0002-1176-3391]{Allison~Youngblood}
\affiliation{NASA Goddard Space Flight Center, Greenbelt, MD 20771, USA}

\author[0000-0002-7119-2543]{Girish M. Duvvuri}
\affiliation{\lab}
\affiliation{Department of Astrophysical and Planetary Sciences, University of Colorado, UCB 389, Boulder, CO 80309, USA}
\affiliation{Center for Astrophysics and Space Astronomy, University of Colorado, 389 UCB, Boulder, CO 80309, USA} 

\author[0000-0002-1912-3057]{DJ~Teal}
\affiliation{NASA Goddard Space Flight Center, Greenbelt, MD 20771, USA}
\affiliation{Department of Astronomy, University of Maryland, College Park, MD 20742, USA}

\author[0000-0001-9207-0564]{P.~Wilson~Cauley}
\affil{\lab}

\author[0000-0002-0726-6480]{Darryl~Z.~Seligman}
\affiliation{Department of the Geophysical Sciences, University of Chicago, Chicago, IL 60637}

\author[0000-0002-5258-6846]{Eric~Gaidos}
\affiliation{Department of Earth Sciences, University of Hawai`i at Manoa, Honolulu, HI 96822, USA}

\author[0000-0002-1337-9051]{Eliza~M.-R.~Kempton}
\affiliation{Department of Astronomy, University of Maryland, College Park, MD 20742, USA}

\author[0000-0003-4733-6532]{Jacob~L.~Bean}
\affiliation{\chicago}

\author[0000-0001-8274-6639]{Hannah Diamond-Lowe}
\affiliation{National Space Institute, Technical University of Denmark, Elektrovej 328, 2800 Kgs.\ Lyngby, Denmark}

\author[0000-0003-4150-841X]{Elisabeth Newton}
\affiliation{Department of Physics and Astronomy, Dartmouth College, Hanover, NH 03755, USA}

\author{Sivan Ginzburg}
\affiliation{Department of Astronomy, University of California at Berkeley, CA 94720-3411, USA}

\author[0000-0002-8864-1667]{Peter Plavchan}
\affiliation{Department of Physics and Astronomy, George Mason University, Fairfax, VA, USA}

\author[0000-0002-8518-9601]{Peter~Gao}
\affiliation{Earth and Planets Laboratory, Carnegie Institution for Science, 5241 Broad Branch Road, NW, Washington, DC 20015, USA}

\author[0000-0002-0298-8089]{Hilke Schlichting}
\affiliation{Department of Earth, Planetary, and Space Sciences, University of California, Los Angeles, CA 90095, USA}

\correspondingauthor{Adina~D.~Feinstein;\\ \twitter{afeinstein20}; \github{afeinstein20}} \email{afeinstein@uchicago.edu}


\begin{abstract}

High energy X-ray and ultraviolet (UV) radiation from young stars impacts planetary atmospheric chemistry and mass loss. The active $\sim 22$~Myr M~dwarf \aumic\ hosts two exoplanets orbiting interior to its debris disk. Therefore, this system provides a unique opportunity to quantify the effects of stellar XUV irradiation on planetary atmospheres as a function of both age and orbital separation. In this paper we present over 5~hours of Far-UV (FUV) observations of \aumic\ taken with the Cosmic Origins Spectrograph (COS; 1070-1360\,\angstrom) on the Hubble Space Telescope (HST). We provide an itemization of $120$ emission features in the HST/COS FUV spectrum and quantify the flux contributions from formation temperatures ranging from $10^4-10^7$\,K. We detect \nflares~flares in the FUV white-light curve with energies ranging from $10^{29} - 10^{31}$~ergs. The majority of the energy in each of these flares is released from the transition region between the chromosphere and the corona. There is a 100$\times$ increase in flux at continuum wavelengths $\lambda < 1100$~\angstrom\, in each flare which may be caused by thermal Bremsstrahlung emission. We calculate that the baseline atmospheric mass-loss rate for \planetb\, is $\sim 10^8$~g~s$^{-1}$, although this rate can be as high as $\sim 10^{14}$~g~s$^{-1}$ during flares with $L_\textrm{flare} \simeq 10^{33}$\,erg\,s$^{-1}$. Finally, we model the transmission spectra for \planetboth\ with a new panchromatic spectrum of \aumic\ and motivate future JWST observations of these planets.
\end{abstract}


\section{Introduction} \label{sec:intro}

Magnetic reconnection provides the energy to accelerate proton and electron beams in the stellar atmosphere and eject stellar plasmas, which result in flare radiation emission and coronal mass ejections (CMEs). Decades of multiwavelength observations of solar and stellar flares, from particularly active stars like AD~Leo, have provided insights into the underlying magnetic reconnection and plasma mechanisms driving these explosive events \citep{brueckner76, poland84, macneice85, mcclymont86, hawley91, porter95, antonova98, Aschwanden00, hawley03, osten05, veronig10, zeng14}. While multiwavelength observational campaigns of flares have been ongoing for decades \citep{hawley03, osten05}, the more recent development of exoplanet atmospheric science has led to a resurgence of these campaigns for exoplanet host stars \citep{macgregor21}. While photometric surveys like Kepler \citep{Borucki10} and all-sky surveys like the Transiting Exoplanet Survey Satellite \citep[TESS;][]{Ricker14} provided crucial insight into the frequency of flares as a function of spectral type \citep{notsu13, davenport14, maehara15, loyd18_flares, guenther19_flares, howard_ward19, Feinstein22}, age \citep{ilin19, feinstein20, ilin21}, and rotation period \citep{doyle18, doyle19, doyle20, howard_ward20,Seligman2022}, detailed spectroscopic studies of stellar flares connect broad-band observations to those observed from the Sun. 

Observations of individual stars with the Extreme Ultraviolet Explorer (EUVE) provided the first EUV flare observations of other stars. This allowed for the opportunity to compare these events with the behavior of the solar corona. \cite{hawley95} observed two flares from the active M3 dwarf AD~Leo simultaneously in optical wavelengths. These data combined with contemporaneous X-ray observations provided strong evidence of the Neupert effect \citep{Neupert1968,Dennis1993}, a model of chromospheric evaporation. They also provided evidence that stellar corona are heated via similar mechanisms believed to be operating in the corona of the Sun. \cite{audard99} completed a two-week long observational campaign with the EUVE of the young solar analogues 47~Cas and EK~Dra. They measured quiescent emission that was 2-3~orders of magnitude greater than that of the Sun, with average plasma temperatures in the corona of $20 - 30\times 10^6$~K. They reported flares with energies of $3 \times 10^{33} - 6 \times 10^{34}$~ergs, which are $1-2$~orders of magnitude \textit{more} energetic than the historic Carrington super-flare event on the Sun \citep{carrington1859}.

Observations from the Measurements of the Ultraviolet Spectral Characteristics of Low-mass Exoplanetary Systems (MUSCLES) survey \citep{france16} demonstrated that the baseline FUV/NUV luminosity increases by a factor of $\sim 100\times$ from early K to late M~dwarfs. Additionally, even optically inactive M~dwarfs exhibit frequent flares in their UV light curves based on the strength of H$\alpha$ and \ion{Ca}{2} \citep{france12, loyd16, diamondlowe21}. A detailed analysis of flares from both inactive and active M~dwarfs from the MUSCLES survey was presented by \cite{loyd18}. This study of Hubble Space Telescope (HST) Cosmic Origins Spectrograph (COS)/Space Telescope Imaging Spectrograph (STIS) light curves revealed that the flares from active stars are an order of magnitude more energetic than inactive stars, but both exhibit the same flare frequency distributions \citep{loyd18, loyd18_flares}, where active stars were defined by a \ion{Ca}{2}\,H\,\&\,K equivalent width (EW) $> 10$\,\angstrom\ and inactive stars had EW$< 2$\,\angstrom.

Time-series photometric missions such as Kepler and TESS led to the discovery of $> 5000$~exoplanets.\footnote{NASA Exoplanet Archive, Update 2022 April 5.} However, only a small hand-full of these planets are $<$~$100$~Myr \citep{david16, mann16, david19_v1298all, david19_v1298b, benatti19, newton19, mann20, rizzuto20, carleo21, martioli21, mann21, bouma22}. Time-series observations in the X-ray/FUV of these host stars have provided insight into the effect of young stellar irradiation on exoplanet atmospheres and may quantify the relative importance of photoevaporation and core-powered mass loss for super-Earths and sub-Neptunes \citep{lopez13, owen17,ginzburg18,gupta19,loyd20,rogers21b}.

AU Microscopii (\aumic) has been the target of extensive observations over the past decade because of its close proximity \citep[$9.72 \pm 0.04$~pc,][]{gaia18, plavchan20}, youth \citep[$22 \pm 3$~Myr,][]{Mamajek14}, and circumstellar debris disk \citep{kalas04, liu04, liu04_follow, metchev05, augereau06, fitzgerald07, Plavchan09}. Two transiting exoplanets orbiting interior to the debris disk were reported recently \citep{plavchan20, martioli21, gilbert22}. As an M~dwarf, \aumic\ may provide crucial insights into planetary formation and atmospheric evolution around the most common stellar type in the galaxy \citep{Henry06}. 

Recent observational and theoretical investigations of \aumic\ have constrained its stellar properties that affect the evolution of the short-period planets, including the magnetic field strength, high energy luminosity \citep{cranmer13}, and flare rate. \cite{Kochukhov20} obtained optical spectroscopic and spectropolarimetric observations to characterize at small and global scales. Zeeman broadening and intensification analysis of Y- and K-band atomic lines yielded a mean field of $\langle B \rangle = 2.2$\,kG. Preliminary Zeeman-Doppler imaging revealed a potential weak, non-axisymmetric magnetic field configuration, with a surface-averaged strength of $\langle B_z \rangle = 88$\,G \citep{Kochukhov20}. This is double the magnetic field strength of $\langle B_z \rangle = 46$\,G presented in \cite{martioli21}. A search for long-term activity cycles in the photosphere of \aumic\ revealed a possible stellar cycle length of $40 - 42$~years, with average brightness changes of $\Delta V = 0.2$\,mag \citep{bondar20}.

The EUVE satellite was used to observe multiple flares on \aumic\ in 1992. \cite{cully93} observed two flares with energies $E_{EUV} = 2\times 10^{33}$ and $3 \times 10^{34}$~ergs at 65-190\,\angstrom\ with estimated temperatures of $3 \times 10^7$\,K. Spectroscopic investigations of these flares revealed that the temporal evoluton of Fe\,\textsc{XX-XXIV} was similar to the photometric light curve. This demonstrated that the hot plasma ($\sim 10^7$\,K) may have experienced rapid expansion and adiabatic cooling \citep{drake94}. The existence of these Fe lines also constrained the differential emission measurement (DEM) models at temperatures between $10^6-10^8$\,K \citep{monsignori94}. Modeling the DEM during different phases of the flares revealed a high-temperature component during the entire event and subsequent decay, with shifts towards higher temperatures at the peak \citep{monsignori96}.

\cite{redfield02} conducted a survey of late-type dwarfs with Far Ultraviolet Spectroscopic Explorer (FUSE) in which \aumic\ was observed flaring twice. Flaring was observed by FUSE in the FUV continuum and in several emission lines including \ion{C}{2} at 1036\,\angstrom, \ion{C}{3} at 977 and 1175\,\angstrom, and \ion{O}{6} at 1032\,\angstrom, which trace formation temperatures from $4.74 \leq \textrm{log}(T[\textrm{K}]) \leq 5.45$. The continuum fluxes were fit with a $\textrm{log}(T[\textrm{K}]) \sim 8.0$ thermal bremsstrahlung profile, in contrast to the blackbody profile more typically used to interpret M dwarf flares at NUV wavelengths \citep{kowalski13}.

The stellar activity of \aumic\ specifically has been characterized recently. The system was also observed in two sectors of TESS data, resulting in two $\sim 27$~day light curves separated by $\sim 1$~year. \citet{gilbert22} reported an average flare rate of $\sim 2$~flares per day with a slight increase in activity after one year. \cite{veronig21} recently conducted an investigation of coronal dimming from coronal mass ejections (CMEs) following flares on \aumic\   with archival XMM-Newton observations. Statistically significant dimming events were seen following three flares in the sample.
 
The activity of \aumic\ provides variable and harsh environments for its planets. \cite{alvaradoGomez22} modeled the space-weather experienced by \planetboth\ in the presence of stellar winds and CMEs. These simulations indicate that \planetboth\ reside inside the sub-Alfvénic region of the stellar wind, with average pressures of $10^2-10^4\times$ the average value experienced by Earth. \cite{alvaradoGomez22} presented simulations of  extreme CMEs in the system with global radial speeds $\sim5\times 10^3-10^4$\,km s$^{-1}$, mass of $\sim 2 \times 10^{18}$\,g, and kinetic energies between $10^{35-36}$\,erg. The CMEs increased the dynamical pressure felt by the planets by $4-6$ orders of magnitude with respect to the steady-state, and could temporarily shift the planetary conditions from sub- to super-Alfvénic.

\begin{figure*}[!bth]
\begin{center}
\includegraphics[width=\textwidth,trim={0.25cm 0 0 0}]{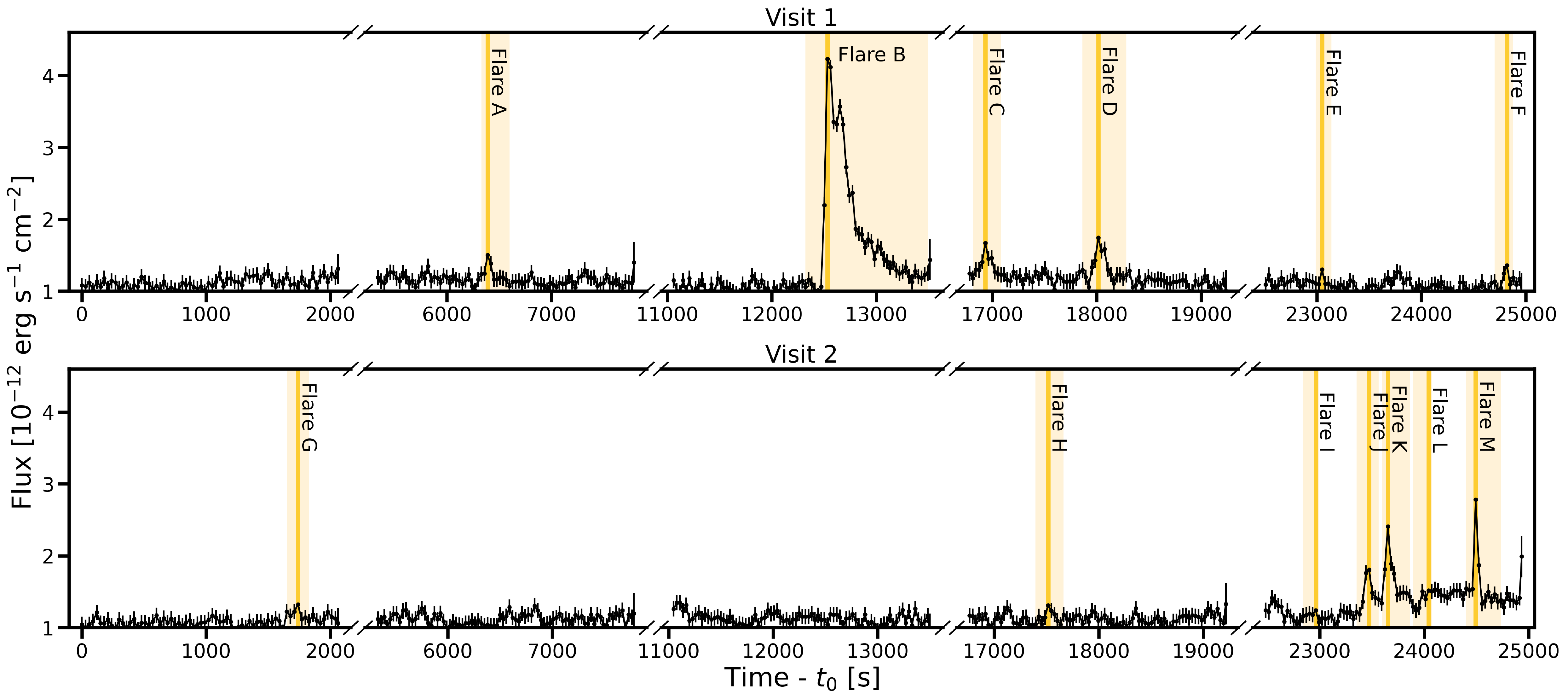}
\caption{Flux-calibrated light curves from two HST/COS visits to \aumic\ across the entire wavelength coverage ($1064 - 1361$\,\angstrom). Time of peak flare events are marked with a vertical orange lines. Highlighted yellow regions are excised for the creation of a clean out-of-flare template spectrum. A total of \nflares\ flares were identified, with one double-peaked flare identified in the third orbit of Visit 1 (Flare B) and 5 flares present in the last orbit of Visit 2 (Flares H-L). We present the parameters for each flare in Table~\ref{tab:flares}. \label{fig:lightcurve}}
\end{center}
\end{figure*}

In this paper, we present time-series observations of \aumic\ with the Hubble Space Telescope (HST) Cosmic Origins Spectrograph (COS) to characterize its flare and quiescent emission in the FUV. This paper is organized as follows. In Section~\ref{sec:obs}, we describe the observations, the creation of light curves, and the identification of flares and spectral emission features. We then describe the properties and morphologies of the flares in Section~\ref{sec:analysis}. We also describe variations of emission line profiles and provide measurements of the continuum flux both in quiescence and during flares.  In Section~\ref{sec:panchromatic}, we provide a panchromatic spectrum of \aumic\ using a combination of models, and current and archival observations. In Section~\ref{sec:planet}, we model the atmospheric mass-loss for \planetb\ and atmospheric retrievals for \planetboth\ with these new FUV observations. We search for evidence of coronal dimming and an affiliated proton beam during the most energetic flare in our sample in Section~\ref{sec:processes}. In Section~\ref{sec:conclusion}, we summarize our results and advocate for future X-ray/FUV observations of flares and JWST observations of \planetboth. We provide Jupyter notebooks for specific sections/figures, which are hyperlinked with the \codeicon\ icon.

\section{Observations \& Reduction}\label{sec:obs}

We observed \aumic\ over two visits with HST/COS under GO program 16164 (PI Cauley). We used the COS G130M grating centered at 1222~\angstrom\ for both visits with $R \sim 12,000-17,000$, following the instrumental configuration used in \cite{froning19}. This configuration provides coverage from approximately 1060-1360~\angstrom\ with a detector gap from 1210-1225~\angstrom, masking the bright \lya\ emission feature to avoid detector saturation. The same COS setting was used for both visits. The visits were executed on 2021 May 28 and 2021 September 23 during transit events of \planetb. The transits of \planetb\ are a separate ongoing analysis and do not impact the flare results presented here. We note the reference start time for Visit 1 is MJD = 59362.148; the reference start time for Visit 2 is MJD = 59480.629.

\subsection{Light Curve Creation \label{subsec:lc} \href{https://github.com/afeinstein20/cos_flares/blob/paper-version/notebooks/extracting_time_tag_data.ipynb}{\codeicon}}

\aumic\ is well known to be active, with flares observed in the far UV \citep{redfield02} and the optical with TESS broadband photometry \citep{gilbert22}. We produced light curves using the \texttt{time-tag} markers available in our HST/COS output files. This mode documents every photon event as a function of time and wavelength, allowing for time-series spectra to be extracted.

In order to categorize the observational data into time bins, we used \texttt{costools}\footnote{\url{https://github.com/spacetelescope/costools}}, which is a set of tools for HST/COS data reduction. The \texttt{costools.splittag.splittag} routine creates time-separated \texttt{corrtag} files for a given number of input seconds. For the primary data reduction, we binned the observations into 30~second exposures. It has been previously established by several sets of authors that the time-resolution can impact measured flare amplitudes and energies \citep{howard22, lin22}. We chose to use 30~second exposures to balance high temporal cadence with sufficiently high signal-to-noise ratio (SNR) per bin. We reduced each new \texttt{corrtag} file using the default processing pipeline of \texttt{calcos}\footnote{\url{https://github.com/spacetelescope/calcos}}, which provides a set of calibration tools for HST/COS. We extracted 1D spectra (\texttt{x1d} spectral data products) from every unbinned \texttt{corrtag} file. 

\begin{figure}[!htb]
\begin{center}
\includegraphics[width=0.46\textwidth,trim={0.25cm 0 0 0}]{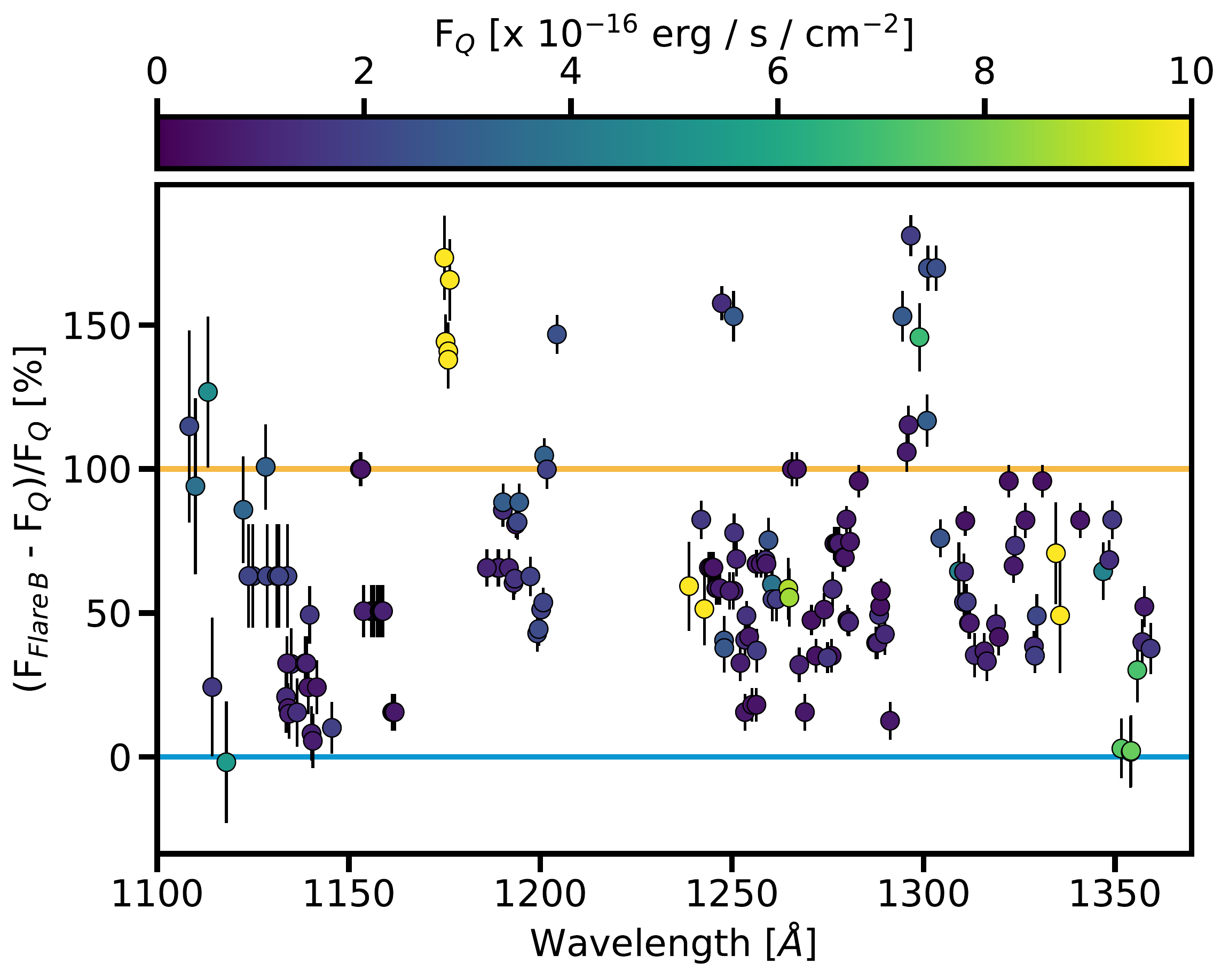}
\caption{Comparison of line flux during Flare~B (see Figure~\ref{fig:lightcurve}) and in quiescence (F$_Q$) for all lines identified in the \aumic\ spectrum. Points are colored by line flux in quiescence. The values and error bars are presented in Table~\ref{tab:linelist}. There is an overall increase in flux for all identified lines during Flare~B as compared to the quiescent state. The blue and yellow lines represent no change and a doubling of flux values. \href{https://github.com/afeinstein20/cos_flares/blob/paper-version/notebooks/big_table.ipynb}{\codeicon} \label{fig:table}}
\end{center}
\end{figure}

There are 402 1D spectra per visit after this reduction, with detector segments a and b for each spectrum covering the full G130M CENWAVE 1222 bandpass. Each 1D spectrum has calculated affiliated errors per each observed wavelength, which we use directly in our error propagation. The wavelength solution per each frame visit is slightly different. To mitigate this issue, we interpolated each 1D extracted spectrum onto the same wavelength grid with a log-uniform dispersion of 0.009\AA\ bin$^{-1}$. Our white light curve (flux summed over our entire wavelength coverage) is shown in Figure~\ref{fig:lightcurve}. We present our light curve in units of seconds, in accordance with previous FUV flare studies \citep[e.g.][]{loyd18, froning19, france20}. We also created light curves of individual spectral features by isolating emission lines in the 30-s cadence 1D spectra for flare identification and analysis.

\subsection{Flare Identification}
Due to the small data set, we identified flares by-eye in each orbit. Flares were identified as large amplitude outliers in the light curves, followed by a decay. Each candidate flare was required to have at least two data points above the noise level of the orbit when it occurred. To identify flares, we searched two separate light curves: the \ion{C}{3} emission line at 1175.59\angstrom\ and the \ion{Si}{3} emission line at 1294.55\,\angstrom\ (Section~\ref{appendix:lcs}; Figure~\ref{fig:id_lc}). This method is consistent with previous studies \citep[e.g.][]{woodgate92}. Flares were more pronounced in the \ion{C}{3} and \ion{Si}{3} light curves, while some smaller flares were not as obvious in the white light curve alone. In total, we identified \nflares\ flares within both visits to \aumic, labeled with with capital letters in Figure~\ref{fig:lightcurve}. Additionally, we highlight all time bins associated with the flare in the yellow shaded region. The peak of the flare is highlighted by a thick vertical line. We consider all remaining time bins outside of the yellow highlighted regions to be attributed to \aumic\ in quiescence. The affiliated spectra are mean-combined to create a quiescent spectrum for \aumic, presented in Figure~\ref{fig:spectrum}.

\subsection{Spectral Line Identification}

The high SNR and high activity level of \aumic\ produces a spectrum rich with emission features, which allow us to create a nearly complete list of present emission features and their measured fluxes, with respect to the databases and published FUV line lists. We searched for known emission features through the CHIANTI atomic database of spectroscopic diagnostics \citep{chianti_main, chianti_v7}, the National Institute of Standards and Technology (NIST) Atomic Spectra Database \citep{nist}, previous HST/STIS observations of \aumic\ \citep{pagano00}, and FUSE observations of \aumic\ \citep{redfield02}. 

We present measured flux values, wavelength/velocity offsets compared to rest wavelengths, and full-width half-maximum (FWHM) values for all identified lines in Table~\ref{tab:linelist}. Measured flux values are presented in units $10^{-15}$\,erg\,s$^{-1}$\,cm$^{-2}$. We measured the line fluxes and FWHMs by assuming each line profile can be modeled by a single Gaussian function convolved with the COS line-spread function. We performed a $\chi^2$ minimization for each line, allowing the mean ($\lambda_\textrm{obs}$; assuming the potential for non-negligible Doppler shifting), line width, and amplitude to vary. We do not explicitly fit for the continuum around each line, but include a term to account for an offset with respect to the continuum. 

This process was completed for the mean quiescent spectrum and for a mean-combined spectrum of all time bins during Flare~B, the largest flare in our sample. We summarize the results of Table~\ref{tab:linelist} in Figure~\ref{fig:table}, by plotting the quiescent vs. Flare~B line fluxes. We find that, on average, bright and faint emission lines increase by a constant value of $\sim 1.5$ during Flare~B. Additionally, we note that the strongest emission features in quiescence do not show the strongest increase in flux during Flare~B, but rather follow a similar trends as all lines identified.

\section{The FUV Flares of \aumic}\label{sec:analysis}

We have identified \nflares\ flares in our sample. Here, we discuss flare parameters as a function of energy, equivalent duration, and wavelength. Additionally, we compare how line profiles and continua change from quiescence during the five most energetic flares in our sample: Flares~B, D, J, K, and M (Figure~\ref{fig:lightcurve}).

\vspace{3mm}
\subsection{Flare Modeling \& Parameters}\label{subsec:modeling}

For broadband optical/IR white-light flares, the typical flare light curve model consists of a sharp Gaussian rise followed by an exponential decay \citep{Walkowicz11, davenport14}. We find that the previous model does not fit the FUV light curves well because the flare peak tends to appear more rounded, rather than discretely peaked. This is likely due to the wavelength dependencies in our light curves. Mainly, the peak of the flares in our FUV light curves are not as sharply peaked as the flares in Kepler and TESS. Instead, we develop a new profile to fit wavelength-dependent flares, which combines the \cite{Davenport16} and \cite{gryciuk17} flare profiles and is similar to the newly developed models by \cite{tovarMendoza22}. Here, we use a skewed Gaussian profile with respect to time, $\textrm{profile}(t)$, convolved with the white-light flare model. The skewed Gaussian takes the form:

\begin{equation}\label{eq:skew}
\begin{split}
 \textrm{profile}(t) \sim \frac{1}{ \sqrt{2 \pi}\omega} \,\bigg(\,e^{- (t - \xi)^2/(2 \omega^2)} \,\bigg)\\\,\times \left[1 + \textrm{erf}\left(\frac{\eta}{\sqrt{2}}\right) \right]
 \end{split}
\end{equation}

\noindent where $\xi$ is the mean time of the distribution, $t$ is time relative to an arbitrary starting point, $\omega$ is a free parameter with units of time that sets the amplitude of the distribution, and erf is the error function. It is important to note that Equation \ref{eq:skew} has units of inverse time, because it will be convolved with respect to time. The parameter $\eta$ is a renormalized and dimensionless proxy for time, and is defined as
\begin{equation}
 \eta = \alpha\left(\frac{t-\xi}{\omega}\right)\,,
\end{equation} 
where $\alpha$ is dimensionless and defines the skew of the distribution. When $\alpha > 0$, the distribution has a steeper rise on the left wing; while a profile with $\alpha < 0$ has a steeper decay on the right wing. For all models, $\alpha > 0$, indicative of a steeper rise.

For completeness, the white-light flare model with respect to time, $\textrm{white-light flare}(t)$, takes the form

\begin{equation}\label{eq:whitelight}
 \textrm{white-light flare}(t) = 
 \begin{cases}
  a \,e^{-(t-t_\textrm{peak})^2/(2r^2)} & t < t_0\\
  a\, e^{-(t-t_\textrm{peak})/d} & t \geq t_0
 \end{cases}
\end{equation}
\noindent where $t_0$ is the time of peak intensity of the flare, $a$ is the amplitude of the flare with units of flux, $r$ is a parameter that sets the slope of the rise of the flare, and $d$ sets the slope of the decay of the flare. The function that we use to fit the flares in this data set is the convolution of Equations \ref{eq:skew} and \ref{eq:whitelight}. After performing the convolution, we calculate best fit parameters by performing a $\chi^2$ minimization, allowing all parameters to freely vary.

\begin{deluxetable}{l r r r r}[!t]
\tabletypesize{\footnotesize}
\tablecaption{Measured Flare Parameters \label{tab:flares}}
\tablehead{
\colhead{Flare} & \colhead{$t_\textrm{peak}$} & \colhead{\textit{E}} & \colhead{ED} & \colhead{$N_\textrm{flares}$}\\
\colhead{} & \colhead{[s]} & \colhead{[10$^{30}$ erg]} & \colhead{[s]} & \colhead{}}
\startdata
A & 6388 & 3.78 $\pm$ 0.14 & 17.4 $\pm$ 9.0 & 1 \\
B & 12531 & 24.1 $\pm$ 0.14 & 688.5 $\pm$ 88.0 & 3\\
C & 16935 & 4.17 $\pm$ 0.32 & 51.6 $\pm$ 20.3 & 1\\
D & 17985 & 3.81 $\pm$ 0.29 & 53.9 $\pm$ 18.1 & 1\\
E & 23049 & 1.19 $\pm$ 0.11 & 1.8 $\pm$ 6.8 & 1\\
F & 24819 & 1.24 $\pm$ 0.11 & 5.9 $\pm$ 6.8 & 1\\
\hline
G & 1740 & 2.42 $\pm$ 0.22 & 6.4 $\pm$ 13.6 & 1\\
H & 17515 & 3.51 $\pm$ 0.32 & 1.0 $\pm$ 20.4 & 1\\
I & 22993 & 2.01 $\pm$ 0.18 & 4.9 $\pm$ 11.4 & 1\\
J & 23473 & 3.50 $\pm$ 0.25 & 60.3 $\pm$ 15.9 & 1\\
K & 23653 & 3.64 $\pm$ 0.22 & 100.8 $\pm$ 13.6 & 1\\
L & 23983 & 1.39 $\pm$ 0.11 & 17.3 $\pm$ 6.8 & 1\\
M & 24493 & 3.53 $\pm$ 0.22 & 92.4 $\pm$ 13.6 & 1\\
\enddata
\tablecomments{The parameter $t_\textrm{peak}$ is the peak time of the flare; $E$ is the flare energy; ED is the flare equivalent duration; $N_{\textrm{flares}}$ denotes the number of flare models combined in the best-fit result. The horizontal line separates flares from Visit 1 and Visit 2. The reference start time for Visit 1 is MJD = 59362.148; the reference start time for Visit 2 is MJD = 59480.629.}
\end{deluxetable}

\begin{figure*}[!ht]
\begin{center}
\includegraphics[width=\textwidth,trim={0.25cm 0 0 0}]{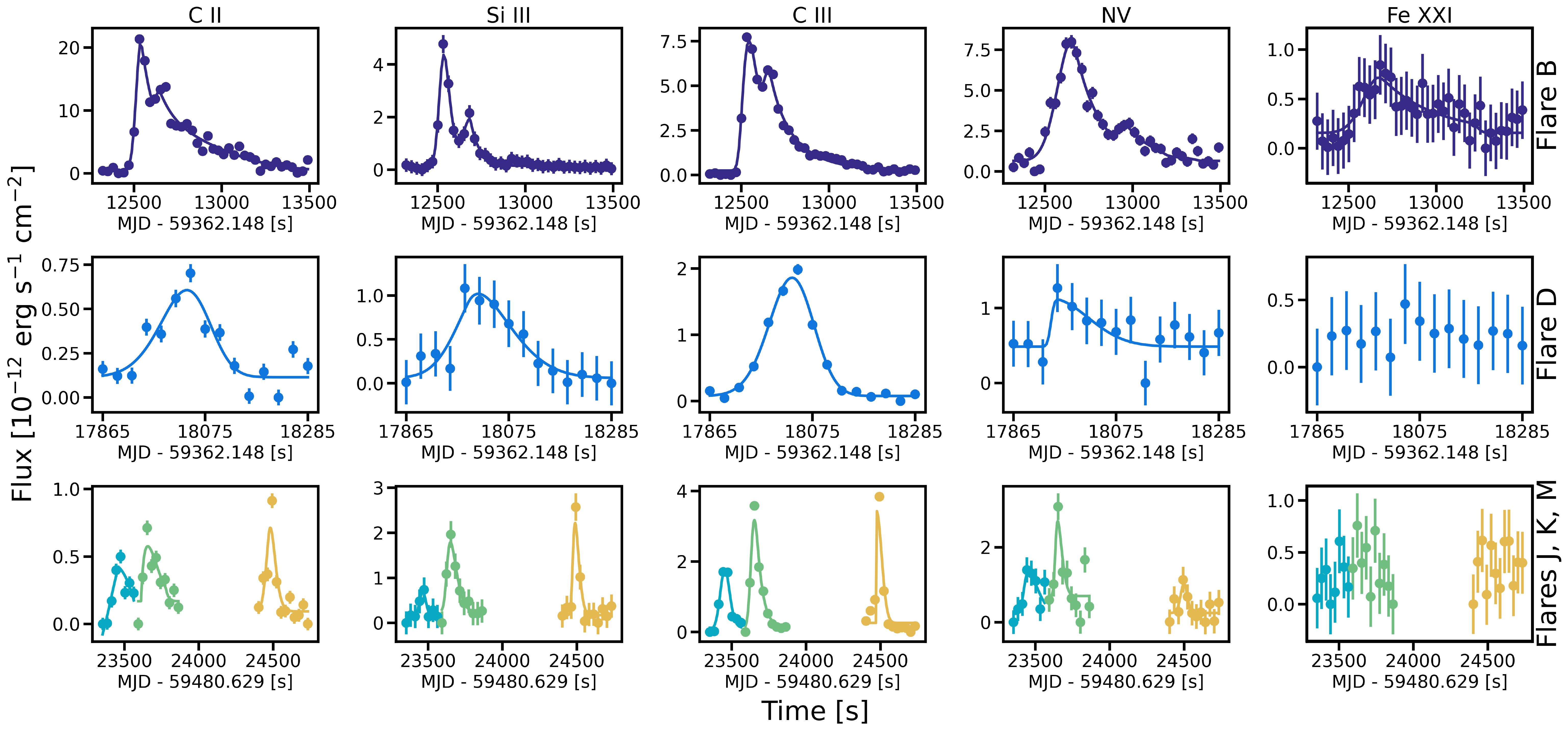}
\caption{A comparison of flares seen in emission lines that originate from different formation temperatures (provided in Table~\ref{tab:ion_flares}), moving from coolest to hottest emission line from left to right. The best-fit model for each flare is over-plotted as a solid line. The first and second rows are light curves for Flare B and D. The third row are light curves for Flares J, K, and M (teal, green, and yellow). We were unable to properly model Flare D, J, K and M in \ion{Fe}{21} due to a lack of obvious flare shape. \label{fig:flare_ions}}
\end{center}
\end{figure*}

Flare~B is considered a complex flare because it exhibits a pronounced double-peaked structure, and required a unique analytic function to fit. It is likely the secondary peak originates from a \textit{sympathetic} flare to the primary. Sympathetic flares are typically defined as spatially correlated with the primary flare, and are often seen on the Sun. Theoretically, the reconnection event causing the primary flare can trigger readjustments of the local magnetic field line topology, which can potentially trigger additional reconnection events \citep{sturrock84, parker88, sturrock90, Lu1991, Schrijver11}. It is not possible to quantify the spatial correlation between these two flaring events deterministically without the ability to spatially resolve \aumic. However, based on the short timescale between the two flaring events compared to the typical flare occurrence rate of the star, it is likely that these two events are correlated. The decay of Flare~B is also slower than typical decay rates in exponential functions. For these reasons, we model Flare~B with three flare profiles: two profiles for the notable double-peaked structure and a third, lower energy flare profile to approximate the prolonged decay. For all other flares, we used a single flare profile.

We computed the absolute flare energies, $E$, following \cite{davenport14, Hawley14, loyd18}, with the equation,

\begin{equation}
 E = 4 \pi D^2 \int_{t_0}^{t_1}\, \bigg(\,F_f(\tau) - F_q(\tau)\,\bigg)\, d\tau
\end{equation}

\noindent where $D$ is the distance to \aumic, $F_f$ is the flux during the flare and $F_q$ is the quiescent flux. For this work, we adopt a distance of $D=9.72$\,pc \citep{gaia18}. The parameters $t_0$ and $t_1$ represent the initial and final times of the flare, and are calculated when the absolute flux returns to the typical continuum value for the star. Additionally, we calculate the equivalent duration, ED, of the flares as 

\begin{equation}
 \textrm{ED} = \int_{t_0}^{t_1} \,\bigg(\,\frac{F_f(\tau) - F_q(\tau)}{F_q(\tau)} \,\bigg)\, d\tau
\end{equation}

We present the measured absolute flare energies, equivalent durations, and time of the flare peak in Table~\ref{tab:flares}. These energies were calculated from the white light flux light curves (Figure~\ref{fig:lightcurve}). We find the flares in our sample range from $E = 1.19\times10^{30} - 2.41\times10^{31}$\,erg and ED $=1\,\,\textrm{to}\,\,689$\,s.

\subsection{Spectroscopic Light Curves}\label{subsec:speclc}

By creating spectroscopic light curves, we can calculate all of the above flare parameters with the goal of understanding the evolution throughout the stellar atmosphere. We created light curves for the five ions presented in Table~\ref{tab:ion_flares}, which are selected to represent a range of formation temperatures from $\textrm{log}_{10} (T_\textrm{form} [$K$]) = 4.5 - 7.1$. In that table, we also present the velocity range [km s$^{-1}$] over which the data were integrated to create the light curves for each spectral line. We present these light curves in Figure~\ref{fig:flare_ions} for Flares~B (top row), D (middle row), J, K, and M (bottom row). We also include the flare model best fits using the analysis presented in the previous subsection, over-plotted as solid lines. Flares~D, J, K, and M were each modeled with a single flare profile. Flare~B exhibited clear evolution in the double-peaked profile, as well as a decay tail that changed shape between emission features. For this reason, we used a two flare profile for Flare~B in the \ion{C}{2} and \ion{N}{5}, a three flare profile for \ion{Si}{3}, and \ion{C}{3} light curves, and a single flare profile in the \ion{Fe}{21} light curve.

\begin{deluxetable}{l r r r r}[!t]
\tabletypesize{\footnotesize}
\tablecaption{Flares in Different Emission Lines \label{tab:ion_flares}}
\tablehead{
\colhead{Ion} & \colhead{$\lambda$ [\angstrom]} & \colhead{Range [km s$^{-1}$]} & \colhead{$\textrm{log}_{10}(T_\textrm{form} [\textrm{K}])$} & \colhead{$N_\textrm{G}$} }
\startdata
\ion{C}{2} & 1335.708 & [-80,60] & $4.5$ & 2\\
\ion{Si}{3} & 1294.55 & [-100,100] & $4.7$ & 3\\
\ion{C}{3} & 1175.59 & [-240,230] & $4.8$ & 7\\
\ion{N}{5} & 1238.79 & [-80,80] & $5.2$ & 2\\
\ion{Fe}{21} & 1354.07 & [-100,100] & $7.1$ & 1\\
\enddata
\tablecomments{$N_\textrm{G}$ is the number of Gaussian profiles combined to fit the given ion emission feature.}
\end{deluxetable}

\subsubsection{Flare Peak Time Offsets in Flare~B}

In this subsection, we investigate if there are any wavelength dependencies in the time at which Flare~B had peak intensity. We complete this analysis for only the doubly-peaked Flare~B, for which we have a sufficiently high SNR to re-reduce the data to shorter time-bins. For this analysis, we follow the procedures presented in Section~\ref{subsec:lc}. We reduce the data from Visit 1 Orbit 3 where Flare~B occurs using time bins of 3\,s. We measure the time offset of each peak with respect to peak time $t_\textrm{WLC}$ of the ``white light" flare (Figure~\ref{fig:lightcurve}; reported in Table~\ref{tab:flares}) to highlight the evolution of both flares. We define the peak time for the primary flare as $t_\textrm{WLC,1}$ and the secondary as $t_\textrm{WLC,2} = t_\textrm{WLC,1} + 120$. We summarize these results in Figure~\ref{fig:energies}.

We find that the primary peaks of \ion{C}{2}, \ion{Si}{3}, and \ion{C}{3} are within $1.5\sigma$ agreement with $t_\textrm{WLC, 1}$. For hotter emission lines, we find primary peak times of $123 \pm 7$~s (\ion{N}{5}) and $151 \pm 41$~s (\ion{Fe}{21}) after $t_\textrm{WLC, 1}$. For the secondary peaks, we find \ion{C}{2}, \ion{Si}{3}, and \ion{C}{3} occur $7 \pm 13$~s, $31 \pm 5$~s, and $9 \pm 5$~s, respectively, after $t_\textrm{WLC, 2}$. As noted above, we did not detect a secondary peak in the \ion{Fe}{21} light curve, likely due to the low SNR. Additionally, at a faster cadence, the $t_\textrm{WLC, 2}$ is ill-constrained in \ion{N}{5}. 

We find no clear trends in peak time offsets with respect to emission lines for Flare~B. While the cooler temperatures, which trace the transition region, peak earlier than in the white-light curve for the primary peak, the opposite is true for the secondary. We note one reason the general shape of the white-light curve deviates from the typical flare profile could be due to emission lines peaking at different times. This could result in a broader peak, rather than a sharp discreteness between the flare rise and decay.

\subsubsection{Energy Contributions}

\begin{figure}[!t]
\begin{center}
\includegraphics[width=0.47\textwidth,trim={0 0 0.25cm 0}]{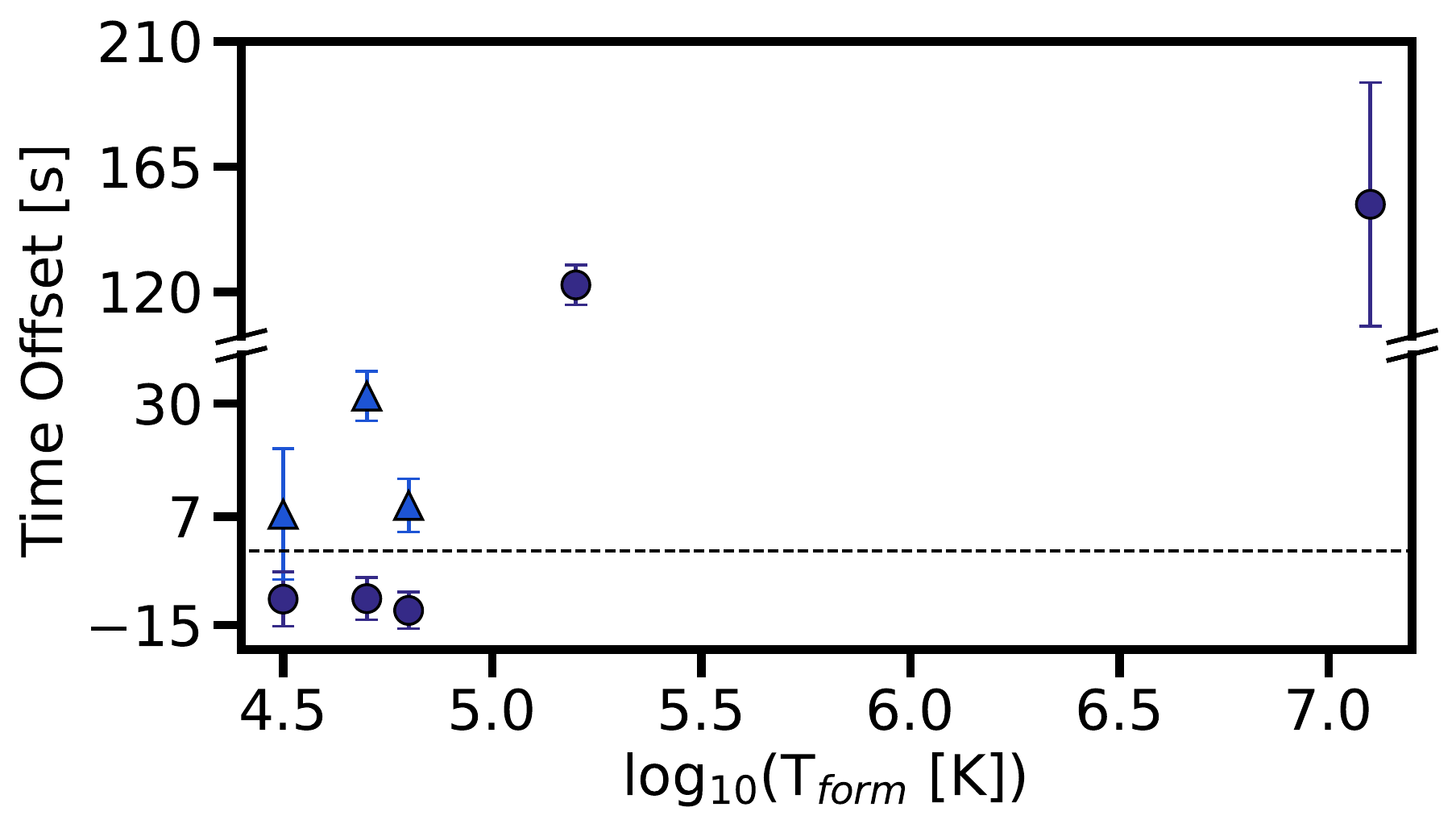}
\includegraphics[width=0.47\textwidth,trim={0.25cm 0 0 0}]{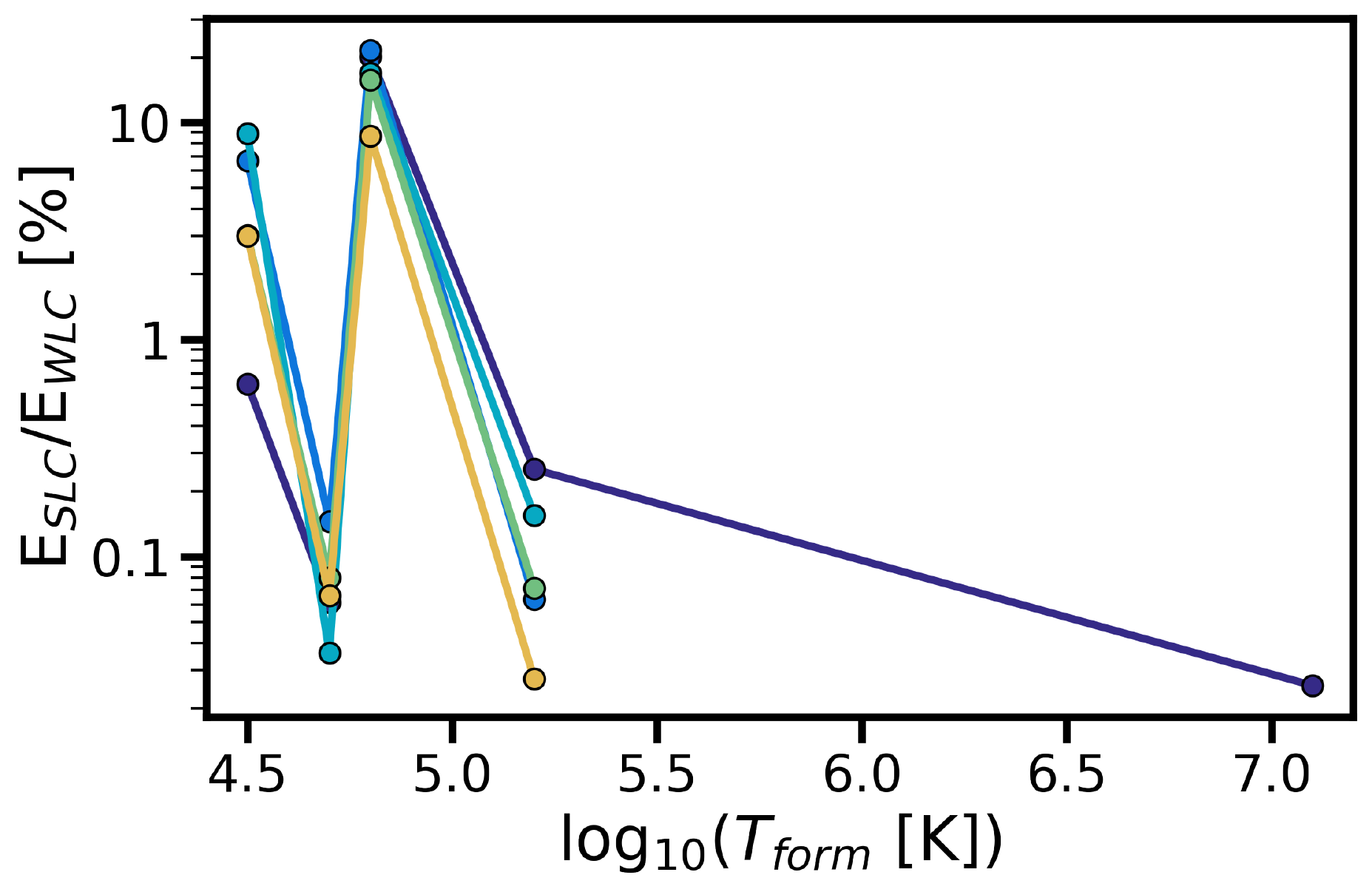}
\caption{Top: A comparison of time offsets for the primary (circles) and secondary (triangles) flare peaks with respect to the peak in the ``white-light" for the complex Flare~B as a function of formation temperature. We plot the zero-point as a horizontal dashed line and note the time offset of the secondary peak in the in ``white light" as a dotted black line. We set the zero-points for the primary and secondary peaks as $t_p = 12531$~s and $t_s = 12651$~s, with respect to the visit start time (MJD = 59362.148). We do not see the secondary peak in \ion{N}{5} and \ion{Fe}{21}. Bottom: A comparison of the measured energies for each flare from the spectroscopic light curves ($E_{SLC}$) compared to the measured white-light energy ($E_{WLC}$). All flares have the strongest measured energy in \ion{C}{3} ($\textrm{log}_{10} (T_\textrm{form} [$K$]) = 4.8$). Flare~B has the highest $E_{SLC}/E_{WLC}$ in \ion{C}{3}, likely due to the increased prominence of the second flare at this wavelength (see Figure~\ref{tab:ion_flares}).
\label{fig:energies}}
\end{center}
\end{figure}

We compare the energies measured in the spectroscopic light curves, $E_{SLC}$, to energies from the full COS G130M band white light curve, $E_{WLC}$. We evaluate the contribution of each emission line to the total white-light flare energy (Figure~\ref{fig:energies}). We find that all flares in our sample follow similar trends in the energies measured from the spectroscopic light curves. Each flare has the largest energy contribution from \ion{C}{3}, followed by \ion{C}{2}. For this analysis, we treat Flare~B as a single flare.

We find the largest contribution from \ion{C}{3} across all flares, where $(E_{SLC}/E_{WLC})_{\textrm{C}\textsc{III}} = 10 - 21$\%. The energies from \ion{C}{2} have the second largest contribution to the total energy, where $(E_{SLC}/E_{WLC})_{\textrm{C}\textsc{II}} = 1 - 7$\%. Interestingly, the weakest contribution of \ion{C}{2} is for the most energetic flare in our observations (Flare~B). We find total contributions of \ion{Si}{3} and \ion{N}{5} to be $(E_{SLC}/E_{WLC})_{\textrm{Si}\textsc{III}} = 0.05 - 0.15$\%, $(E_{SLC}/E_{WLC})_{\textrm{N}\textsc{V}} = 0.07 - 0.28$\%, respectively, and, for Flare~B, $(E_{SLC}/E_{WLC})_{\textrm{Fe}\textsc{XXI}} = 0.03$\%. 

These trends are suggestive of the energy from the flare being deposited deeper in the upper chromosphere and transition region, while coronal heating is negligibly affected for these observed flares. Simultaneous observations of Flare~B in the X-ray would have provided a better constraint on the high energy contribution to the total output, and how the flare affects hotter plasma in the stellar atmosphere.

\begin{figure*}[!ht]
\includegraphics[width=\textwidth,trim={0.25cm 0 0 0}]{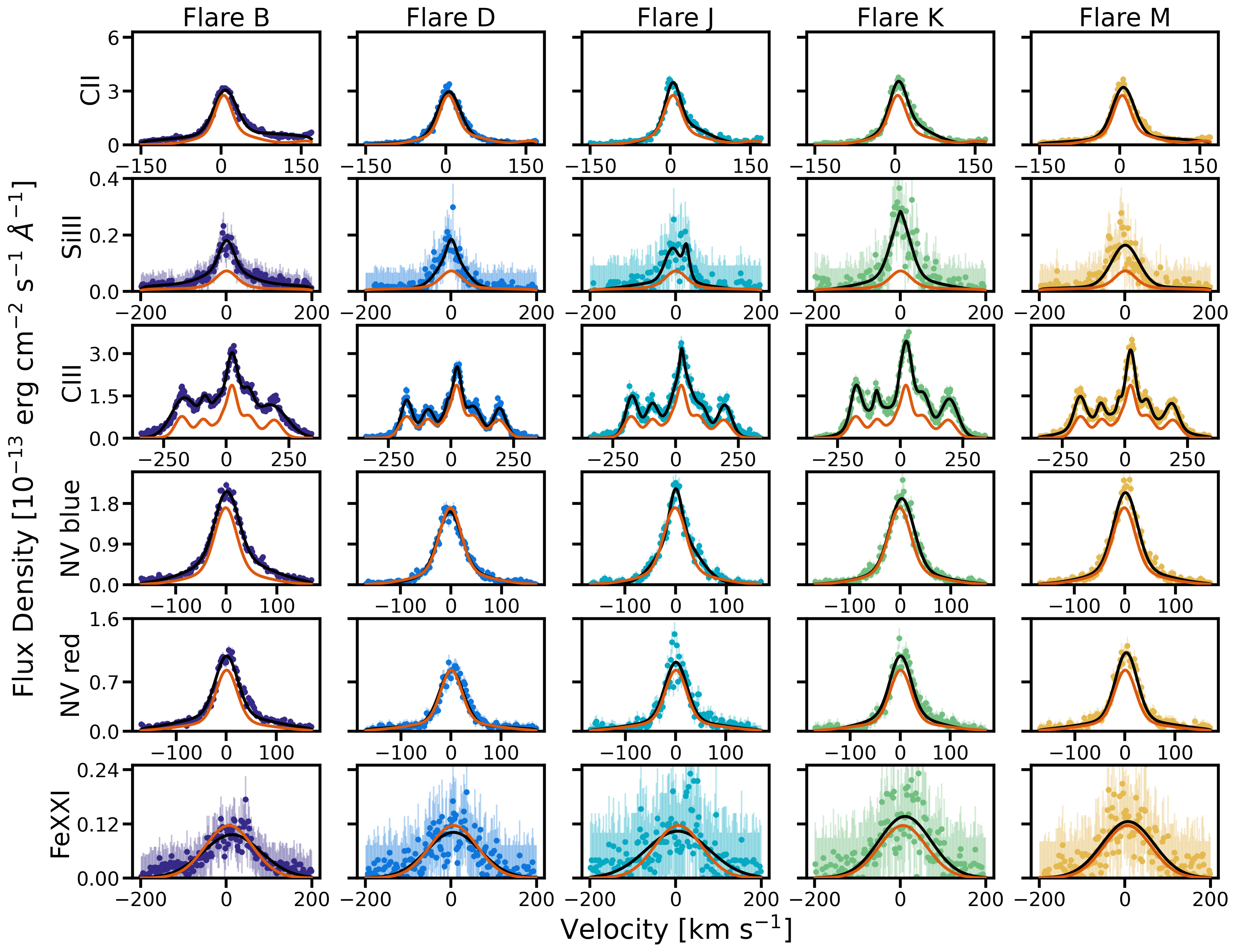}
\caption{A comparison of line profiles in quiescence compared to Flares B, D, J, K, and M (left to right). The best-fit quiescent line profile is plotted in orange; the best-fit in-flare line profile is plotted in black, with the data plotted in color. All line profiles were fit with a multi-Gaussian model, where the exact number of Gaussians in each model is presented in Table~\ref{tab:ion_flares}. We find that for Flares B, J, K, and M each ion exhibits a bulk flux increase. Flare D poses the only exception to this, where there is little change in the profiles of \ion{C}{2}, \ion{N}{5}, and \ion{Fe}{21}. In \ion{Si}{3}, Flares B, J, K, and M all show a bulk increase in the blue side of the line center. In \ion{N}{5} doublet, Flares B, J, K, and M all exhibit additional flux in the peak and red-wing of both the blue and red components. \href{https://github.com/afeinstein20/cos_flares/blob/paper-version/notebooks/line_analysis.ipynb}{\codeicon} \label{fig:lineprofiles}}
\end{figure*}

\subsection{Line Profiles}

In addition to modeling differences in flare morphologies and measuring differences in energies, we evaluate changes in the line profiles of the emission features. In this subsection, we evaluate this for every feature listed in Table~\ref{tab:ion_flares} in quiescence and during Flares~B, D, J, K, and M. Each of the profiles are presented in Figure~\ref{fig:lineprofiles}. Each is modeled with multiple Gaussian profiles convolved with the Line Spread Function (LSF) of COS. The best fitting model is calculated using a $\chi^2$ best fit between the data and the model using \texttt{lmfit} \citep{lmfit}. The number of Gaussians, $N_G$, used for each line profile is listed in Table~\ref{tab:ion_flares} and the best-fit model is plotted as a solid black line in Figure~\ref{fig:lineprofiles}. For comparison, the best-fit model for the quiescent line profile is plotted as a solid orange line.

We find that, across all flares explored, the FWHM of the best-fit line profiles increases between quiescence and in-flare. We report the following changes in FWHM between quiescence and Flares~B, D, J, K, and M: $10\,\,\textrm{to}\,\,35$\% for \ion{C}{2}, $130\,\,\textrm{to}\,\,293$\% for \ion{Si}{3}, $37\,\,\textrm{to}\,\,94$\% for \ion{C}{3}, $-3 \,\,\textrm{to}\,\, 22$\% for the blue component of \ion{N}{5}, $2 \,\,\textrm{to}\,\, 29$\% for red component of \ion{N}{5}, and $-19 \,\,\textrm{to}\,\, 17$\% for \ion{Fe}{21}. We note that in the blue component of \ion{N}{5}, only Flare~D was found to have a smaller FWHM during the flare than in quiescence. Additionally, Flares~B, D, and J all have smaller FWHM in \ion{Fe}{21} during the flares.

On average, there is additional redshifted emissions (e.g. \ion{C}{2} and \ion{N}{5}) and the peak of the line is redshifted during the flares by up to 15\,km\,s$^{-1}$. Additional redshifted emission (30-200\,km\,s$^{-1}$) has been found during other M~dwarf flares \citep{redfield02, hawley03, loyd18, loyd18_flares}. This feature is believed to trace material flowing downward toward the stellar photosphere.

\subsection{Comparison of Continua}\label{subsec:continua}

We investigate our spectra for changes in the quiescent and flare continua to measure the differences in best-fit blackbody temperatures, which is often assumed to be $\sim 9000$\,K \citep{kretzschmar11}. We defined regions of the spectra without any emission features as the continuum \citep[following][]{froning19,france20}. The continuum extends across the entire wavelength coverage of G130M. We provide the specific wavelength regions of the continuum in Appendix~\ref{appendix:continuum}.

\begin{figure}[!t]
\begin{center}
\includegraphics[width=0.47\textwidth,trim={0.25cm 0 0 0}]{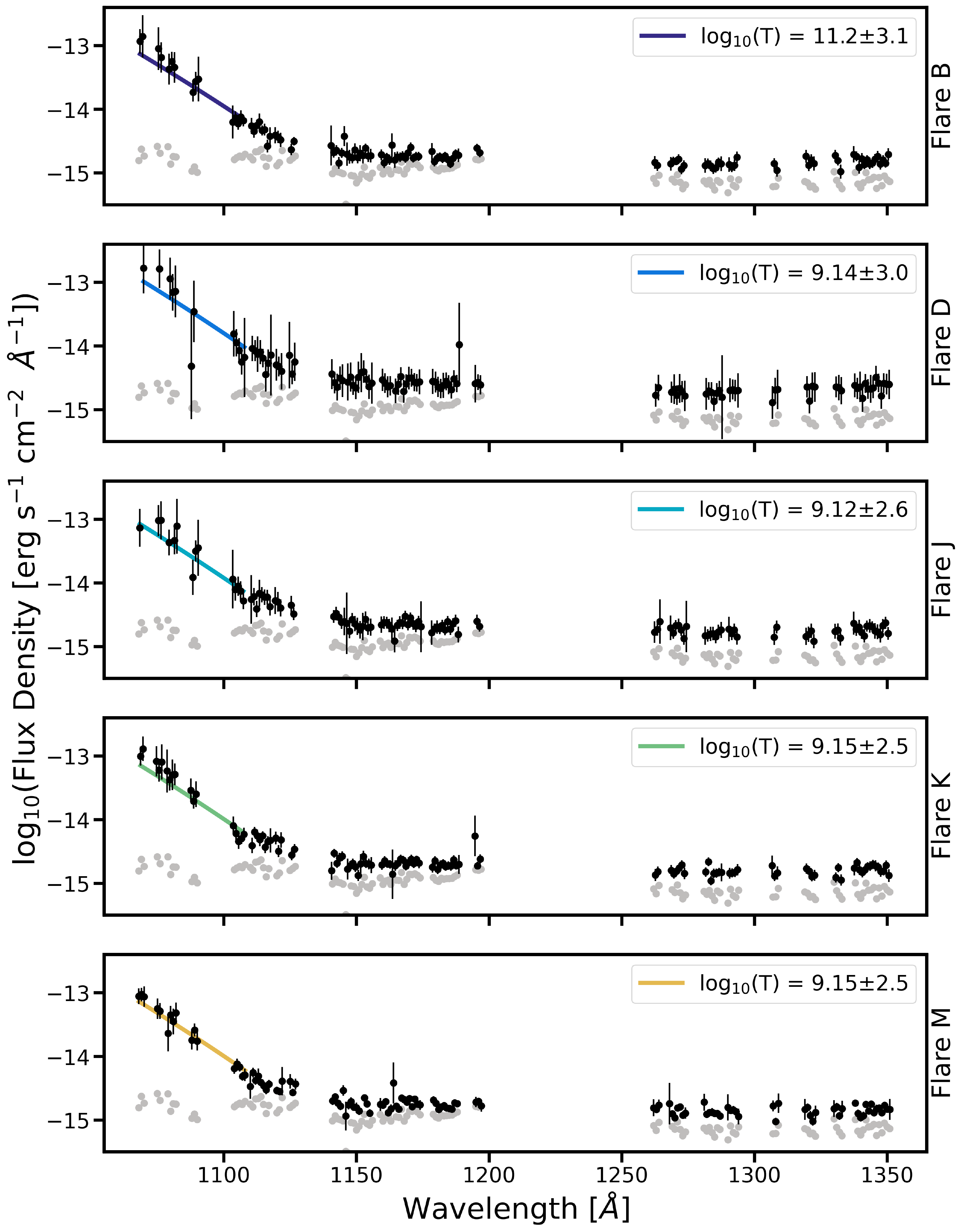}
\caption{The continuum of the mean spectrum for the quiescent state (gray points) and Flares B, D, J, K, and M (black points per each sub-panel). The continuum was visually identified by inspecting regions of the spectrum lacking emission features. For the purposes of this calculation, the spectrum is sub-divided into 1\angstrom\ bins. We fit an additional thermal bremsstrahlung profile (colored lines) to the continuum of the flare data, as there is an obvious rise at $\lambda \leq 1100$~\angstrom. The resulting best-fit temperature for the thermal bremsstrahlung profile is presented in each sub-panel. \href{https://github.com/afeinstein20/cos_flares/blob/paper-version/notebooks/blackbody_fits.ipynb}{\codeicon} \label{fig:bbs}}
\end{center}
\end{figure}

We present our continua for the quiescent state, and Flares B, D, J, K, and M in Figure~\ref{fig:bbs}. The continuum points are subdivided into 1~\angstrom\ bins. To characterize the temperature of the continuum, we fit an ideal blackbody at $\lambda \geq 1120$~\angstrom. In the quiescent state, we find a best-fit blackbody of 16,300~$\pm$~500~K; during the flares, we find the best-fit blackbody ranges from 14,900~--~15,700~K. This is consistent with the blackbody emission from an FUV superflare observed by \cite{loyd18_flares} on another young M~dwarf and towards the upper end of continua emission seen during 20 M~dwarf flares \citep{kowalski13}. Given that these high temperatures are present in the quiescent state of this cool star, it is unclear whether or not a blackbody is the appropriate model to fit to these data.

Thermal bremsstrahlung is a principal emission mechanism for the Soft X-ray emission observed in solar flares \citep{shibata96, warren18, McTiernan19}. We fit for both the temperature and electron number density in a thermal bremsstrahlung profile at $\lambda \leq 1120$~\angstrom. We found temperatures of $9.1 \leq \textrm{log}_{10}(T) \leq 11.2$ best-fit the continua increases seen during Flares~B, D, J, K, and M (models plotted in Figure~\ref{fig:bbs}). However, we note these fits converge only for electron number densities of $\sim 10^{22}$\,cm$^{-3}$, which is not representative of the stellar chromosphere. Therefore, thermal bremsstrahlung cannot be solely responsible for this feature, and it is unclear what other mechanisms may be contributing to this FUV excess.

We visually inspected the COS images to determine if this was the result of an overall count rate increase on a portion of the COS segment b detector. We found the count rate increase is limited to within the spectral trace, lending confidence to an astrophysical origin of this signal. To investigate further, we attempted to fit the slope with a blackbody function. Specifically, we attempted to fit only the \textit{slope}, rather than the overall flux density values. However, we find that the blackbody fit fails to converge, as the function cannot accommodate the two orders of magnitude change in flux density over $\sim 30$\,\angstrom.

Observations of \aumic\ with FUSE found a similar increase in flux during two flares \citep{redfield02}. The best-fit temperature for a bremsstrahlung profile for these data was calculated to be $\textrm{log}(T) \sim 8.0$. We compare the blue end of our continua ($\lambda \leq 1120$~\angstrom) to the continua presented in \cite{redfield02}. We find the continuum during Flare B has a steeper flux decrease between $1066-1115$~\angstrom, decreasing by $\Delta$F~$= 0.877 \times 10^{-13}$~erg~cm$^{-2}$~s$^{-1}$, while the bigger flare in \cite{redfield02} shows a decrease of $\Delta$F~$= 0.003 \times 10^{-13}$~erg~cm$^{-2}$~s$^{-1}$ from $955-1104$~\angstrom.

While we find tentative evidence of thermal bremsstrahlung emission during these flares, we note that this emission mechanism implies a large continuum enhancement at EUV wavelengths that would be several orders of magnitude brighter than bound-bound emission lines and recombination continuua in the EUV (see Section~\ref{subsec:dem}). Such continuum enhancement was never observed with EUVE during flares from \aumic\ \citep{cully93, monsignori96} or other stars. Similar flare continuua have not been identified in other M~dwarf FUV flare observations \citep{loyd18_flares, froning19}. Regardless of the physical mechanism producing the FUV continuum rise, it likely extends into the EUV bandpass where it will contribute to the atmospheric escape of AU Mic b and AU Mic c following stellar flares.

\section{A Panchromatic Spectrum of \aumic}\label{sec:panchromatic}

\aumic\ has been observed across nearly all wavelengths. We performed a systematic search of archival data to construct a panchromatic, quiescent spectrum of this exquisite system -- depicted in Figure~\ref{fig:panchromatic}. In this section, we describe each data set and models employed for unobserved wavelength regimes in order to create our panchromatic spectrum. We do not include archival EUVE observations due to the low SNR of the spectra, and they do not cover the entire EUV wavelength range.

\subsection{XMM-Newton Observations}

Several observations of \aumic\ are available on the XMM-Newotn Science Archive. We used data from Obs.ID 0822740301 (PI Kowalski) which span a wavelength range of $\lambda = 4.02 - 40$~\angstrom. \aumic\ was observed from 2018 Oct 10 to Oct 12. This program includes time-series X-ray observations which captured multiple flares. The majority of these data were taken in quiescence. For the remainder of this analysis, we assume that the median spectrum is representative of \aumic\ in quiescence.

\subsection{FUSE Observations}

Observations of \aumic\ were obtained by FUSE \citep{moos00, sahnow00} as part of the ``Cool Stars Spectral Survey" program. \aumic\ was observed on 2000 Aug 26 and 2001 Oct 10 over 905-1187~\angstrom\ in the time-tagged mode, allowing for the separation of in-flare vs. out-of-flare spectra. The details of these observations are presented in \cite{redfield02}, who identified two temporally-resolved flaring events. For the panchromatic spectrum presented in this paper, we removed the in-flare spectra and use the average of all of the remaining quiescent spectra in the FUSE observations. 

\subsection{Ly$-\alpha$ Reconstruction}

\lya\ is a key driver in planetary atmospheric photochemistry and must be included in the panchromatic spectrum. However, \lya\ is masked in our HST/COS observations for detector safety (see \citealt{osten18}). We note that \cite{wood21} reconstructed a \lya\ profile for \aumic\ in the interest of understanding imprints from the stellar wind on \lya\ observations. In the construction of the SED presented here, we chose to use the model \lya\ profile presented in \cite{flagg22}.

\cite{flagg22} used archival STIS observations of \aumic\ (1999-09-06; \citealt{pagano00}) to detect \lya\ in quiescence and reconstruct its profile following the methods of \cite{youngblood16}. The wings of the model \lya\ profile are fainter than measured by COS (see Figure~\ref{fig:spectrum}) because of (i) stellar variability between 1999 and 2021 and/or (ii) differences between the STIS and COS flux calibrations. To account for these differences, we uniformly scale the model profile by a factor of 18 to match the COS wings.

\subsection{NUV Observations}

We retrieved 20 archival observations of \aumic\ with the International Ultraviolet Explorer (IUE) in the near-ultraviolet (NUV) covering $1750 - 3450$~\angstrom. Observations were taken from 16 January 1986 to 29 July 1991 as part of programs HC078 \citep[PI Butler;][]{butler86} and MC111 (PI Byrne). All observations were taken with low dispersion, large aperture, and exposure times of 1200~s. There were no flares reported in either programs \citep{quin93, maran94}, which we verified visually. We used the median of all NUV observations for the baseline quiescence value. 

\subsection{Optical Spectrum}

In this subsection, we describe how we reconstructed the quiescent optical spectrum of \aumic\ using two methods. In the first method, we obtained 19 publicly available spectra of \aumic\ taken from 2019-2021 from the HARPS-N (3789-6912~\angstrom) data archive. In order to obtain \textit{only} the quiescent spectrum, we removed any spectra with dramatic changes in H$\alpha$ -- indicative of a flaring events/periods of increased stellar activity. Specifically, we removed spectra with (i) strong H$\alpha$ emission and (ii) asymmetric profiles caused by an increase the blue wing of the H$\alpha$ flux \citep{maehara21}. This analysis resulted in a data set for the quiescent state which included 11 spectra. We then used the average of this combined data set to produce the optical component of our panchromatic spectrum. We verified that the numerical values of each individual wavelength bin $\delta\lambda$ was consistent with the mean value within $2\sigma\, \forall \delta \lambda$ in the panchromatic spectra. 

In the second method, we extended the range of our optical spectra from that which was observed using a PHOENIX stellar model \citep{husser13}. These models are high-resolution synthetic spectra generated assuming local thermodynamic equilibrium in the stellar atmosphere. Specifically, we selected a model with an effective temperature of $T_\textrm{eff}=3700$~K, surface gravity of log($g$)~=~4.5, and a solar-type metallicity ([M/H]~=~0). The PHOENIX model in our panchromatic spectrum spans from the end of the HARPS-N spectrum to 5~$\mu$m, the wavelength cutoff used in the MUSCLES high-level science products \citep{loyd16}.

\subsection{Differential Emission Measure}\label{subsec:dem}

In this subsection, we describe the methodology by which we estimate the UV and X-ray flux at wavelength regimes not covered by archival observations. Specifically, we calculate the differential emission measure (DEM) which can be used to estimate unobservable EUV flux. Typically, these wavelengths are difficult to observe because of (i) the faintness of the target and/or (ii) photon obscuration from the interstellar medium. We use only the HST/COS observations as inputs to the DEM. W do not use the archival XMM-Newton and EUVE data in our fits. The reason for this is that the scaling between quiescence and during flares for non-simultaneous data most likely does not accurately represent the most recent observations. In order to calculate the DEM model, we follow the methods presented in Section 3 in \cite{duvvuri21}.

This implementation of the modeling takes into account the following procedures:
\begin{itemize}
 \item It assumes a constant electron pressure across the stellar atmosphere.
 \item It incorporates the width of the line emissivity function while fitting the DEM model.
 \item It groups together ions of the same species and calculates the total emissivity across all spectral lines.
 \item It accounts for multiplets.
 \item It assumes that the systematic uncertainty, $s$, can be inferred during fitting by parameterizing it as a fraction $s$ of the DEM predicted flux.
 \item It calculates a free-bound and two-photon continuum component from H and He.
\end{itemize}
The inclusion of the continuum is an update to the methods published in \cite{duvvuri21}.\footnote{\url{https://github.com/gmduvvuri/dem\_euv}} 

\begin{figure}[!tb]
\begin{center}
\includegraphics[width=0.47\textwidth,trim={0.25cm 0 0 0}]{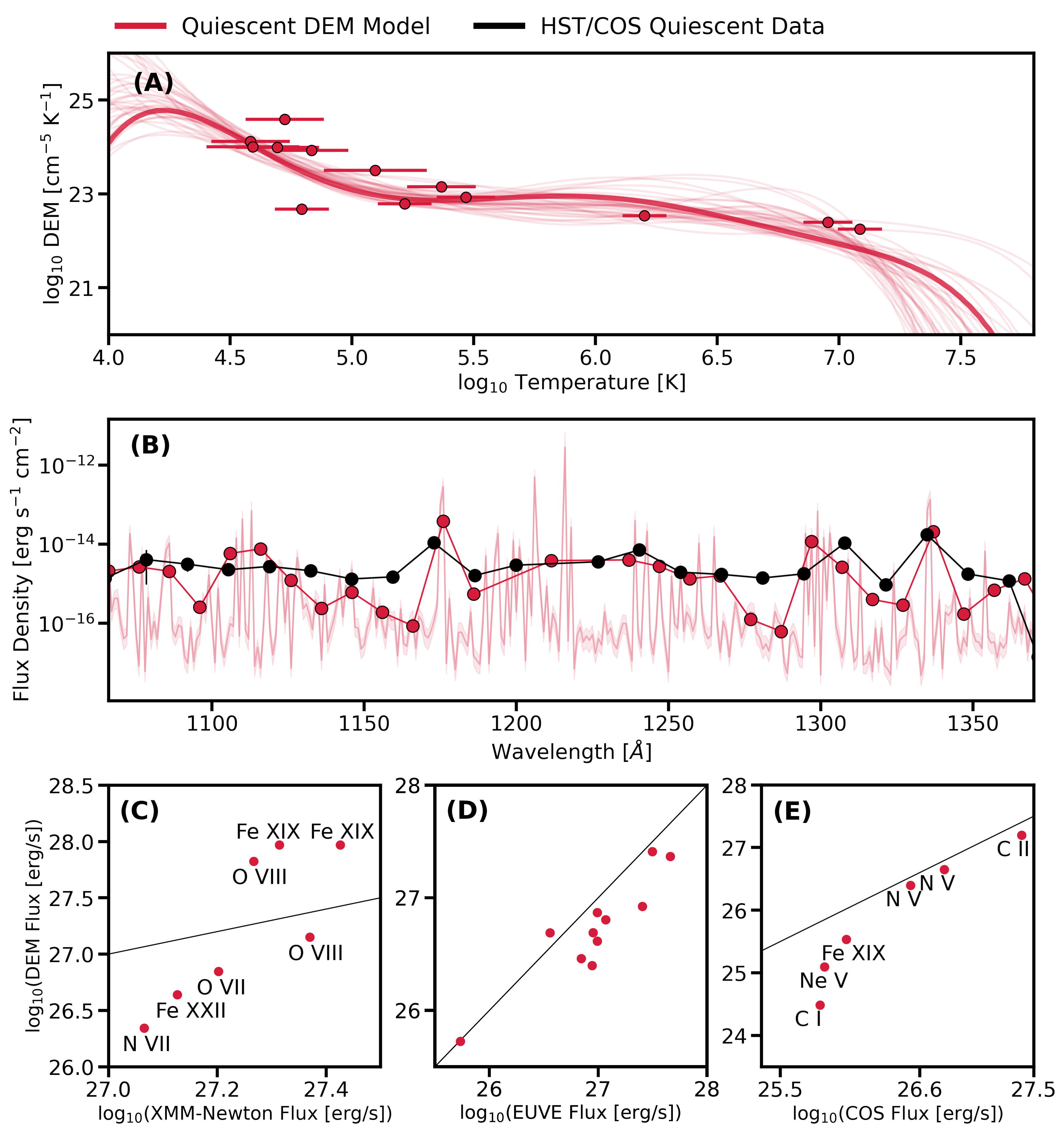}
\caption{A set of Differential Emission Measurement (DEM) models and diagnostic plots. For all of the plots, red lines and symbols represent the DEM models, while black lines and symbols represent the HST/COS data presented in this paper. Panel (A) shows the DEM models of the COS FUV emission lines of \aumic\ in quiescence. The average DEM model is shown in the thick line, with individually measured DEM values. The thin lines represent 50~random draws from the models fit with \texttt{emcee}. Panel (B) shows the integrated flux in bins of 10\angstrom\ from the DEM output (red) spectra compared to the HST/COS data (black). The \lya\ line is masked in these bins. The unbinned spectrum is plotted as the pink line with shaded $1\sigma$ errors. In Panel (C) we show a comparison of line fluxes from XMM-Newton observations of \aumic\ to the DEM modeled spectra. In Panel (D) we show a comparison of line fluxes from EUVE observations \citep{delzanna02} of \aumic\ to the DEM modeled spectra. These data are for \ion{Fe}{9} - \ion{Fe}{24}. In Panel (E) shows a comparison of line fluxes from the presented HST/COS observations to line fluxes. The solid lines in both Panels (C, D, and E) represent a 1-to-1 relationship in the flux. \href{https://github.com/afeinstein20/cos_flares/blob/paper-version/notebooks/dem_figure.ipynb}{\codeicon} \label{fig:dem}}
\end{center}
\end{figure}

\begin{figure*}[!ht]
\begin{center}
\includegraphics[width=\textwidth,trim={0.25cm 0 0 0}]{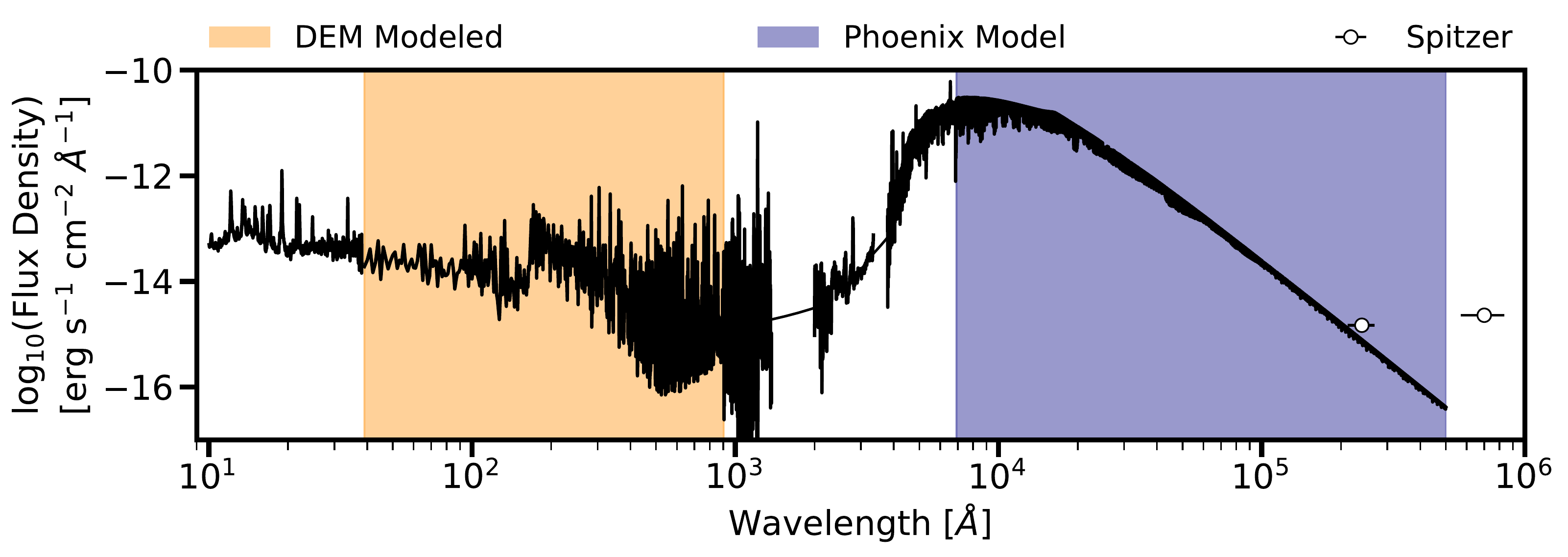}
\caption{A panchromatic spectrum for \aumic\ in its quiescent state. The spectra are comprised of archival observations of \aumic\ with XMM-Newton (10\,-\,39\,\angstrom), FUSE (900\,-\,1181\,\angstrom), HST/COS (this work; 1064\,-\,1372\,\angstrom), IUE (2000\,-\,3347\,\angstrom), and HARPS-N (3789\,-\,6912\,\angstrom). To fill in gaps in data coverage, we utilize a DEM synthetic spectrum (40-900\,\angstrom), a linear interpolation (1372\,-\,2000\,\angstrom\ and 3334\,-\,3782\,\angstrom), and a PHOENIX synthetic generated stellar atmosphere model (6912\,-\,$2\times10^5$\,\angstrom). For \lya\ (1211\,-\,1220\,\angstrom), we use the reconstructed profile from \cite{flagg22}. We present Spitzer 24 and 70\,$\mu$m color-corrected detections of \aumic\ (white points) for completeness \citep{Plavchan09}. We do not correct the PHOENIX spectrum for the infrared excess from the debris disk. \href{https://github.com/afeinstein20/cos_flares/blob/paper-version/notebooks/panchromatic_spectrum.ipynb}{\codeicon} \label{fig:panchromatic}}
\end{center}
\end{figure*}

To determine the DEM parameters, we use the measured fluxes for lines marked with an asterisk (*) presented in Table~\ref{tab:linelist}. These lines are known to have emissivities dominated by gases with temperatures of $10^{4-8}$\,K, and mostly neglects neutral lines. Additionally, these selected lines are not blended with any other known emission line of comparable emissivity and have strong enough emissivities that the line identification routines are reliable. These lines are single resolved lines or multiplets which fit within a 1\,\angstrom\ bin, which ensures all relevant emissivity is added into the model properly. Further selection criteria for reliable emission lines are described in \cite{duvvuri21}.

We use \texttt{CHIANTI 10.0.1} \citep{chianti_main, delzanna21} to calculate the emissivity functions for all transitions in the \texttt{CHIANTI} database. We assumed solar coronal abundances \citep{schmelz12} and calculated the emissivity functions from $4 \leq \textrm{log}_{10}(T) \leq 8$. We fit a Chebyshev polynomial to to reproduce the measured line fluxes. We then ran a Markov Chain Monte-Carlo (MCMC) fit for the coefficients of the polynomial and the estimated systematic uncertainty fraction, $s$. For the MCMC, we used 50 walkers and 1200 steps. We visually verified that the walkers were sufficiently randomized after the first 400~steps, which were subsequently discarded. We present our DEM measurements and functions of \aumic\ in quiescence in Figure~\ref{fig:dem}. The synthetic spectra generated from the DEM model are shown in red and our HST/COS data are shown in black.

We present several diagnostics to validate our DEM model. First, we subdivide the FUV flux estimated by the DEM models into 10\,\angstrom\ bins and compare to the observations (Figure~\ref{fig:dem}, panel B). As is evident in the figure, there is good overall agreement between the estimated and observed flux. We note the model does under-predict largely line-free spectral regions by two orders of magnitude. We attribute these differences to the additional quiescent FUV continuum emission that we describe in Section~\ref{subsec:continua}. 

We explicitly did \textit{not} model \lya\ to compare with the reconstruction. The reason for this choice is that the \lya\ is not formed under the physical conditions applicable to the DEM method. Therefore, it would be unphysical to include the DEM-estimated \lya. We further validate these methods by evaluating the line flux from specific X-ray, EUV, and FUV emission lines (Figure~\ref{fig:dem}, panels C, D, \& E). We find the X-ray, EUV, FUV estimated flux values from our DEM model are generally consistent with the XMM-Newton, EUVE \citep{delzanna02}, and HST/COS observations.

The DEM presented here is generally consistent with the one presented in \cite{delzanna02}, as also highlighted in Figure~\ref{fig:dem} Panel (D). However, there are minor differences in the overall shape of this DEM. The \cite{delzanna02} DEM model of \aumic\ has a well-constrained peak at log$_{10}(T) = 6.1$ from EUVE measurements and at log$_{10}(T) = 6.9$ from FUSE and STIS observations. The differences at the high temperature regime (log$_{10}(T) >6.5$) may be caused by (i) different observational data sets used to generate each of the DEM models (ii) the differences between the polynomial fit or (iii) more precise laboratory measurements of atomic spectral data since 2002.

\section{Implications for AU~Mic~b and c} \label{sec:planet}

There is no consensus as to whether stellar flares contribute to or detract from the habitability of a planet. For M~dwarf planets specifically, flares may serve as sources of visible light for photosynthesis \citep{airapetian16, mullan18}. They also deliver UV photons needed to initiate prebiotic chemistry \citep{ranjan17, rimmer18}. However, bursts of high-energy radiation and stellar energetic particles (SEPs) from flares can remove the atmosphere of a planet and alter its chemical composition \citep{airapetian20}.

The recent detection of two transiting planets around AU Mic \citep{plavchan20, martioli21} has resulted in significant observational follow-up of the system for planetary and stellar characterization. Several campaigns have ensued to measure the masses of \planetboth\ via radial velocities, yielding masses from $M_b = 11 - 20 M_\oplus$ and $M_c \leq 22 M_\oplus $ \citep{cale21, klein21, martioli21, zicher22}. \planetboth\ are near a 9:4 mean-motion resonance, which may produce transit-timing variations \citep{martioli21} and a second means to measure the masses of these planets. Transmission spectroscopy in the optical/near-infrared has proven challenging given the strength of the stellar activity of \aumic\ \citep{palle20, hirano20}.

Young and short period transiting planets are believed to host more extended atmospheres due to the high levels of stellar irradiation \citep[e.g.][]{lammer03, owen19}. This effect is more prominent for stars $< 100$~Myr because the overall stellar XUV flux is higher. Moreover, young and newly formed planets are likely still undergoing contraction. Therefore the characterization of the the atmosphere of \planetboth\ are important benchmarks for understanding planetary evolution.

In this section we discuss two potential implications of our observations for the atmospheres of \planetb\ and \planetc. In Subsection~\ref{subsubsec:massloss}, we investigate the effects of the measured high-energy luminosity and flares of \aumic\ on mass-loss due to photoevaporation. Next, in Subsection~\ref{subsubsec:obssig}, we produce synthetic mid-IR transmission spectra for \planetb\ and \planetc\ using our new panchromatic quiescent spectrum (Figure~\ref{fig:panchromatic}).

\subsection{Flare-Driven Thermal Mass Loss}\label{subsubsec:massloss}

Recently, several sets of authors have focused on flare-driven atmospheric removal during the first 1\,Gyr of the lifetime of a planet. \cite{garcia-sage17} modeled the EUV-driven proton and O$^+$ escape from an Earth-like planet around Proxima Centauri. They speculated that very large flares could increase the ionization fraction at low altitudes. This would indirectly enhance atmospheric escape. They also demonstrated that very energetic flares produce enhanced rates of hydrogen photoevaporation. \cite{feinstein20} modeled H$_2$ dominated atmospheres in the presence of flares on young G~stars following the methods of \cite{owen17, owen20}. They found the inclusion of flares could result in $4-5\%$ more atmospheric mass-loss than without accounting for flares. \cite{NevesRibeiro22} accounted for the X-ray and UV (XUV) contribution of flare flux in atmospheric escape from Earth-like planets around M~dwarfs. The XUV flux from flares produced surface water loss for planets with mass $M_p = 5 M_\oplus$ in their simulations. 

However, the effects of radiation from frequent high-energy flares on short period, young planets has not been fully investigated. Here, we follow methods similar to \cite{feinstein20} to evaluate the effects of flares on the photoevaporation-driving mass-loss of \planetboth. We use the modified energy-limited escape methodology presented in \cite[][Equation 17]{owen17} and \cite{owen20}:

\begin{equation}\label{equation:mdot}
 \dot{M} = \eta \, \bigg(\, \frac{R_p^3 L_{\rm HE}}{4 a^2 G M_{\textrm{Core}}} \, \bigg)\, .
\end{equation}

\noindent In \ref{equation:mdot}, $\eta$ is the dimensionless heating efficiency, $R_p$ is the planetary radius, $L_{\rm HE}$ is the integrated high-energy luminosity of the star from the X-ray through the UV, and $M_{\textrm{Core}}$ is the mass of the core. We make the following assumptions in our model:
\begin{itemize}
 \item The mass loss efficiency is $\eta = 0.15$ \citep{Kubyshkina18}.
 \item The iron mass fraction in the core is $X_{\textrm{iron}} = 1/3$. 
 \item The ice mass fraction in the core is $X_{\textrm{ice}} = 0$.
 \item We adopt the planetary parameters presented in \cite{martioli21}.
\end{itemize}

To evaluate whether these equations can be used to accurately describe \planetboth, we calculate their Jeans parameter. The Jeans parameter, $\lambda_J$, is a quantification of the relative importance of thermal energy and self gravity. It can be calculated using the equation,
\begin{equation}
 \lambda_J = \,\bigg(\, \frac{G\,M_p\,m_H}{k_B \,T_\textrm{exo} \, R_\textrm{exo} } \,\bigg)\,,
\end{equation}
where $M_p$ is the mass of the planet, $m_H$ is the mass of a hydrogen atom, $k_B$ is the Boltzmann constant, $T_\textrm{exo}$ is the temperature at the exobase, and $R_\textrm{exo}$ is the radius of the exobase. 
We calculate the Jeans parameter, $\lambda_J$, for \planetboth\ using the lower mass estimates from \cite{zicher22}. We calculate $\lambda_{J,b} = 1.8$ and $\lambda_{J,c} = 12.8$ using the following two scaled relationships:

\begin{equation}
 \lambda_{J,b} = 1.8 \, \bigg(\, \frac{M_p}{6.7 M_\oplus} \,\bigg)\, \bigg(\, \frac{1500 K} {T_\textrm{exo}} \,\bigg)\, \bigg(\, \frac{14 R_\oplus} {R_\textrm{exo}} \,\bigg)\, ,
\end{equation}
and
\begin{equation}
 \lambda_{J,c} = 12.8 \, \bigg(\, \frac{M_p}{15.5 M_\oplus} \,\bigg)\, \bigg(\, \frac{1000 K} {T_\textrm{exo}} \,\bigg)\, \bigg(\, \frac{6 R_\oplus} {R_\textrm{exo}} \,\bigg)\, .
\end{equation}
This validates the use of the hydrodynamic escape equation for \planetb, but not for \planetc\ \citep{volkov11, gronoff20}. 

We inject flares using the average flare rate found in the observations presented in this paper ($\sim 2.5$\,hour$^{-1}$). We calculate the mass-loss rate for a variety of core masses (5, 8, and 10 $M_\oplus$). We also calculate the timescale over which flares may have an impact on the atmospheric masses. We do this by running our model over three scenarios: (1) no flares are present, (2) flares are present for the first 200~Myr, and (3) flares are present for the first 1~Gyr. 

\begin{figure}[!b]
\begin{center}
\includegraphics[width=0.47\textwidth,trim={0.25cm 0 0 0}]{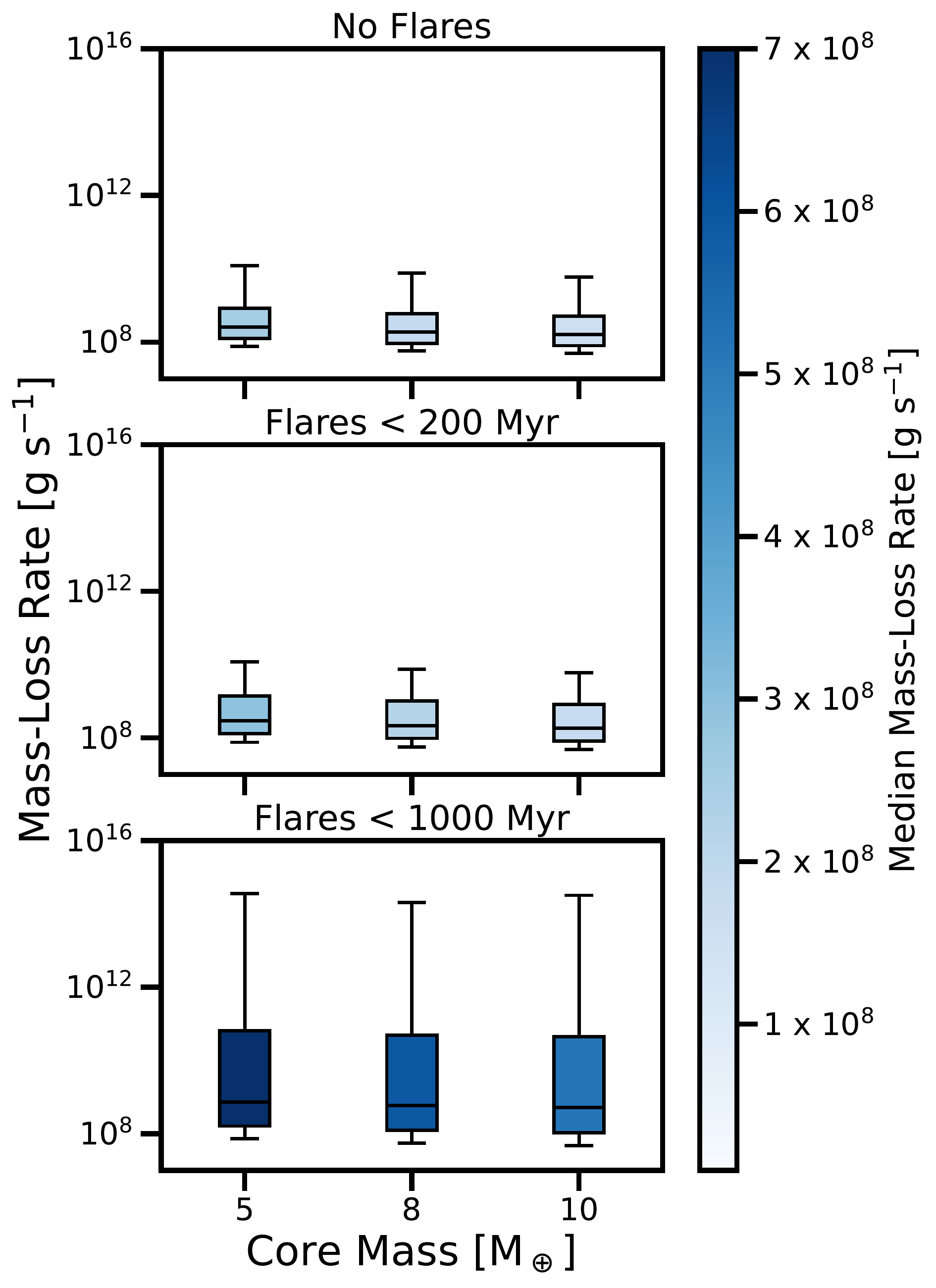}
\caption{Comparison of photoevaporation-driven atmospheric mass loss for \planetb. We run our calculations under three different flare evolution scenarios: (Top) No flares present (Middle) Persistent flares during the first 200\,Myr (Bottom) Persistent flares during the first 1000\,Myr. The box represents the first quartile ($Q_1$; $25^\textrm{th}$ percentile), the median ($50^\textrm{th}$ percentile), and the third quartile ($Q_3$; $75^\textrm{th}$ percentile). The whiskers mark the interquartile (IQR\,$=Q_3-Q_1$), where the lower limit is defined as $Q_1 - 1.5\times$\,IQR and the upper limit is defined as $Q3+1.5\times$\,IQR. The largest spread is seen in the calculation where we inject flares for the first $1000$\,Myr. The boxes are colored by the median mass-loss rate in [g\,s$^{-1}$]. Super-flares ($L_\textrm{flare} > 10^{33}$\,erg\,s$^{-1}$) can boost mass-loss by up to five orders of magnitude. \href{https://github.com/afeinstein20/cos_flares/blob/paper-version/notebooks/flare_mass_loss.ipynb}{\codeicon} \label{fig:massloss}}
\end{center}
\end{figure}

We adopt a quiescent luminosity $L_{HE} = 2.71 \times 10^{29}$\,erg\,s$^{-1}$. We calculate this value by integrating the DEM modeled spectrum over $1 \leq \lambda [\AA] \leq 1100$, for which the DEM is reliable. To simulate flares, we adopt a transient $L_{HE}$ that is equivalent to that of a flare. In the ideal scenario, we would estimate $L_{HE, \textrm{flare}}$ by drawing flares from a fit to the observed flare-frequency distribution. However, since there are only a relatively small sample of flares observed for the system, we adopt a flare-frequency distribution slope of $\alpha = -1.1$~flares\,day$^{-1}$. This value is consistent with that observed on low mass ($M/M_\odot \leq 0.3$) stars \citep{Feinstein22}. We do not account for thermal bremsstrahlung or the continuum in our calculation of $L_{HE}$. Our resulting mass-loss rates are presented in Figure~\ref{fig:massloss}.

First, we calculate the mass loss rates without any stellar flares. We find that the median mass-loss rate for \planetb\ across all assumed core masses ranges from $1.6$ to $2.5 \times 10^8$\,g\,s$^{-1}$ in the case of no flares. These calculations are consistent with the upper limit set by \cite{hirano20} using the metastable infrared \ion{He}{1} triplet with NIRSPEC/Keck-\textsc{ii}. When we include flares for the first 200~Myr, we find no significant change in the time-averaged mass loss rate, with minimal increases of up to $1.5 \times$ the no-flare baseline. When we include flares for the first 1000~Myr, we find the time-averaged mass loss rate increases by $3 \times$ the no-flare baseline. 

Additionally, we can evaluate instantaneous mass-loss in the presence of super-flares ($L_\textrm{flare} \geq 10^{33}$\,erg\,s$^{-1}$) with these simulations. In each simulation, we identify the most energetic flare to be $L_\textrm{flare} \approx 4 \times 10^{33}$\,erg\,s$^{-1}$. We find the instantaneous mass-loss increases by six orders of magnitude, up to $\dot{M} \sim 10^{14}$\,g\,s$^{-1}$, relative to the no-flare baseline. Given the high duty cycle of \aumic, where $1/6^\textrm{th}$ of its time (assuming an average ED$_\textrm{flare} = 5$\,minute) is spent flaring, this could indicate that flares are the dominant source of atmospheric mass removal. There are still many open questions which need to be addressed to claim the previous statement. It is unclear what the response time of the atmosphere would be to being hit by a flare. Additionally, if the atmosphere cannot respond quickly enough to the instantaneous change, then the atmospheric mass loss rate would not increase. This raises the question of if \lya\ or \ion{He}{1} at 1083.3\,nm transits could be variable in depth/shape.

In context with other planets, \lya\ transits have revealed mass loss rates from $10^8$\,g\,s$^{-1}$ \citep{bourrier16} to $2.20\eta × 10^{10}$\,g\,s$^{-1}$ for GJ~436\,b ($R_p = 4.2 R_\oplus$; $M_\star = 0.812 M_\odot$; \citealt{addison19}). No \lya\ transit was detected for the 750\,Myr planet K2-25\,b \citep{rockcliffe21}. Variable \lya\ transits have been observed for HD~189733\,b ($R_p = 12.54 R_\oplus$; $M_\star = 0.45 M_\odot$). \citep{levacelier2010} observed three transits of HD~189733\,b with HST/STIS and constrained the mass loss rate to $10^{9-11}$\,g\,s$^{-1}$. \cite{lecavelier12} observed additional transits one year apart and found changes in the \lya\ transit depth of $\approx 15$\%. They note an X-ray flare was observed 8~hours prior to the deeper transit, although the correlation between events is unclear. \cite{hazra22} recently simulated transits in the presence of flares and CMEs for HD~189733\,b. Although the 3D radiation hydrodynamic simulations revealed transit depth increases by 25\% for flares-alone and a factor of 4 for CMEs, neither models were able to reproduce the deep transit of HD~189733\,b post flare.

\subsubsection{Model Limitations}

The above calculation only accounts for thermal processes, which does not encapsulate all processes which can contribute to atmospheric mass loss. The thermal escape calculation in Section~\ref{subsubsec:massloss} can play a major role if \planetb\ is a low-gravity planet with an atmosphere dominated by light atoms, while non-thermal processes can dominate under a range of planetary configurations \citep{lundin07} and have no mass preference. Many of these processes are understood from our own Solar System. From studies of Mars, \cite{chassefiere07} and \cite{lundin07} defines four primary non-thermal processes.

The first process is photochemical escape, where recombination or charge-charge exchange as the result of stellar ions excites neutral atoms to $v > v_\textrm{escape}$ \citep{lammer96, chassefiere04}. The second process is ion sputtering, which is the result of coronal ions impacting atmospheric neutral particles, resulting in ejection \citep{jakosky94, leblanc02}. The third process is ionospheric escape driven by energy and momentum exchange between the solar wind and planetary atmospheres \citep{moore99}. The fourth process is ionospheric ion pickup, which is the result of both the electric and magnetic fields from the solar wind interacting with and removing ionospheric ions \citep{luhmann91, dubinin06}.

These non-thermal processes can be driven by coronal mass ejections \citep{Lammer07} or interactions with the stellar wind, plasma escaping from the star embedded in the magnetic field \citep{alfven50, parker58}. \cite{cohen22} recently simulated the stellar environment for \planetb\ in the presence of stellar winds, estimated from magnetograms derived by \cite{klein21}, and CMEs. These simulations suggest a potential strong stripping of magnetospheric material from the planet. However, the simulations are unable to quantify the rate of mass-loss per CME interaction.

Magnetic shielding from the presence of a magnetosphere may reduce the efficiency of atmospheric removal \citep{lundin07}. However, the magnetic field strength of \planetboth\ is currently unknown and is essential for understanding these questions. Overall, these non-thermal processes may result in a more significant contribution to atmospheric mass loss than photoevaporation alone. A full calculation coupling both thermal and non-thermal time-dependent processes would need to be completed to fully understand the effect of flares on atmospheric removal, which is outside of the scope of this paper.

\subsection{Observational Signatures}\label{subsubsec:obssig}

\begin{figure*}[!ht]
\begin{center}
\includegraphics[width=\textwidth,trim={0.25cm 0 0 0}]{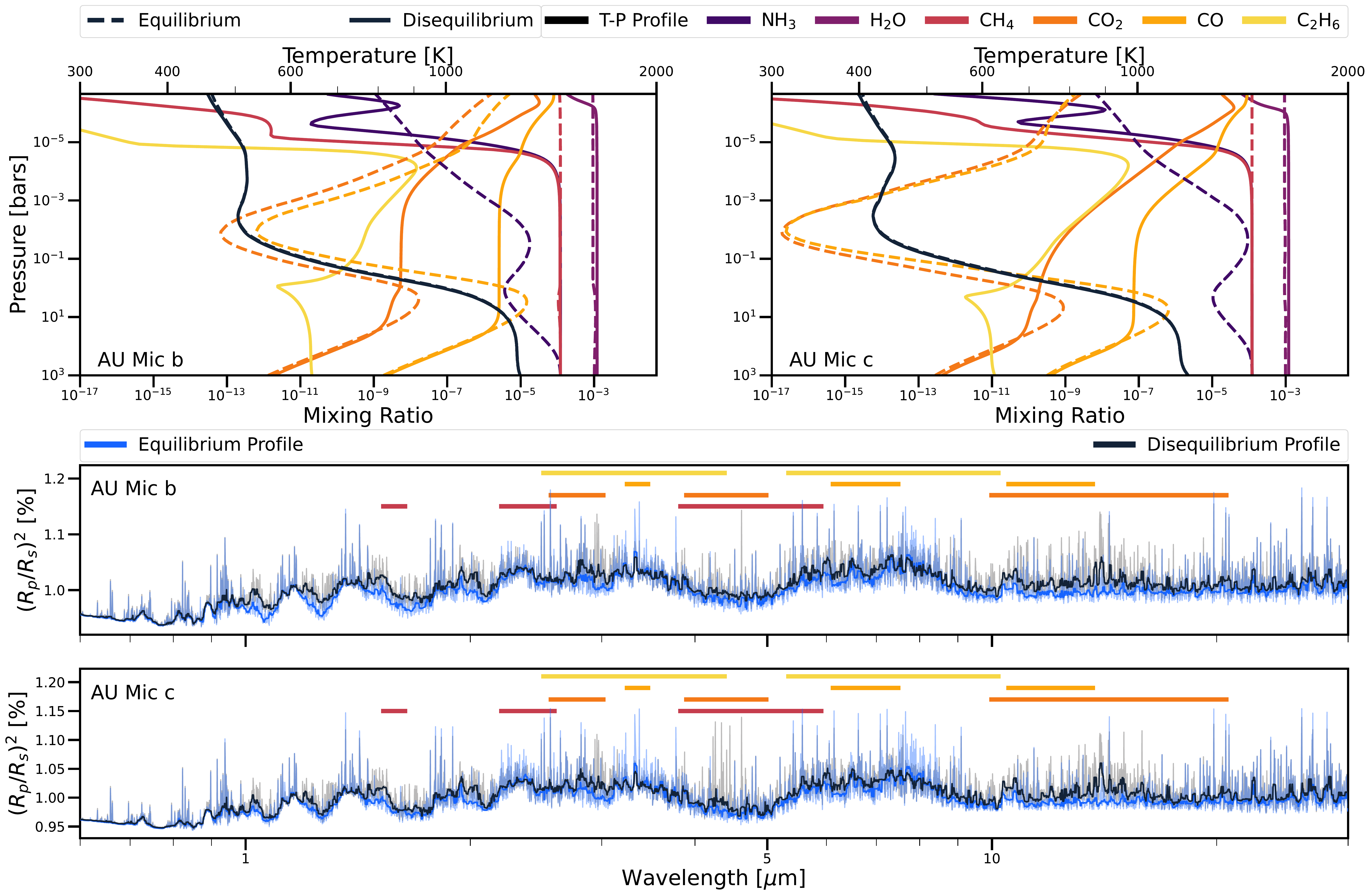}
\caption{Photochemical models for \planetboth. Temperature-pressure profiles (black lines) and mixing ratios (colored) for \planetb\ (top left) and \planetc\ (top right). We model the planets in equilibrium (dashed) and disequilibrium (solid). Normalized transmission spectra as observed from $0.6 - 12 \mu$m for \planetb\ (middle row) and \planetc\ (bottom row). Dominant CH$_4$, CO$_2$, CO, and C$_2$H$_6$ are labeled in the normalized transmission spectra. These models follow the methods presented in \cite{teal22} and are evaluated with \aumic\ in quiescence. \href{https://github.com/afeinstein20/cos_flares/blob/paper-version/notebooks/tp_profile.ipynb}{\codeicon} \label{fig:transmission}}
\end{center}
\end{figure*}

Stellar activity also affects the chemical composition of planetary atmospheres via photochemistry and atmospheric escape. \cite{chen21} presented chemistry-climate model simulations that explored the effects of G, K, and M~dwarf flares on the atmospheres of rocky planets. They demonstrated that the time-averaged flares and accompanying energetic particles can significantly alter the chemical composition of the atmospheres. The global NO and OH increased by an order of magnitude, while the global O$_3$ decreased by less than an order of magnitude after 300 days of post-flare evolution in the atmospheres of planets around M~dwarfs.

The atmospheres of \planetb\ and \planetc\ could be pristine tracers of their primordial atmospheres, although they may have experienced metal enrichment by accreting comets \citep{seligman2022_comets}. Nevertheless, measuring elemental/compound abundances can provide constraints as to where these planets originally formed within the protoplanetary disk \citep{oberg11}. The chemistry and long-term stability depends sensitively on the XUV irradiance of the host star \citep{tian08, Johnstone15}. Here, we model transmission spectra of \planetb\ and \planetc\ in quiescence using the panchromatic spectrum presented in Figure~\ref{fig:panchromatic}. We note that \planetboth\ have the highest Transmission Spectroscopy Metrics ($\geq 350$, \citealt{kempton18}) of all known young transiting exoplanets, making these planets priority targets for future JWST observations.

Here we summarize the methods used for this calculation. We run the \texttt{Atmos} 1-D photochemical model for solar composition atmospheres of \planetboth. We use a recently updated version of \texttt{Atmos} which is appropriate for atmospheres of sub-Neptune (i.e.\ hydrogen-rich) composition, as described in \citet{harman22}. This updated version includes the addition of reactions for nitrogen-bearing species and the hydrocarbon haze prescription from \citet{arney16, arney17}. The temperature-pressure profiles used were computed with the \texttt{HELIOS} radiative-convective equilibrium radiative transfer code \citep{malik17, malik19} for \planetboth~analog planets with 500~K and 600~K equilibrium temperatures, respectively. The photochemical modeling was conducted using the stellar input spectrum for \aumic~from Figure~\ref{fig:panchromatic}, scaled such that the top-of-atmosphere flux corresponds to the orbital distances of each planet as reported by \cite{martioli21}. We then run the resulting atmospheric abundance profiles from the photochemical modeling through the \texttt{Exo-Transmit} radiative transfer code \citep{kempton17} to predict the transmission spectra for both planets. For this calculation, we followed the methods presented in detail in \citet{teal22}.

Figure \ref{fig:transmission} shows two \texttt{Atmos} disequilibrium (black) and two \texttt{FastChem} \citep{stock18} equilibrium (blue) models for \planetboth, as well as affiliated mixing ratios and temperature-pressure profiles. Both cases use the same temperature pressure profiles, since we do not account for the feedback of disequilibrium chemistry on the thermal structure of the atmosphere.

Although we include hydrocarbon haze formation pathways in each of our \texttt{Atmos} models, neither of our atmospheres form significant amounts of photochemical haze. JWST will be able to obtain in-transit spectra from $0.5-28\,\mu$m. It is unclear what level of contamination from stellar activity will be present in these data \citep{zellem17, rackham18}. Any of the instruments on JWST can be used to measure the transit depth and The higher resolution of NIRSPEC compared to NIRISS would make this an ideal instrument to observe H$_2$O and CO$_2$ at $\lambda < 5.3\, \mu$m. Additionally, MIRI could be used to look at H$_2$O, CH$_4$, and O$_3$ at $\lambda > 5\,\mu$m. 

While any of its instruments can be used to measure the transit depth and distinguish between the equilibrium and disequilibrium models, the primary differences for \planetboth\ lie at $\lambda < 3\, \mu$m. For \planetb\ at $1 \leq \lambda \leq 2\,\mu$m, we estimate differences in transit depths of $\approx 200$\,ppm between the two presented models. While for \planetc, we estimate differences of $\approx 70$\,ppm. For \planetboth\ at $3 \leq \lambda \leq 5\, \mu$m, we estimate differences transit depths of $\approx 160$\,ppm. All values predicted by these models are above the estimated noise floor for JWST \citep{matsuo2019, schlawin2020, schlawin2021}.

\cite{teal22} identified that uncertainties in the UV continuum of exoplanet host stars are the primary drivers of uncertainties of photochemical models for hazy exoplanets. With the addition of \aumic's continuum in our panchromatic spectrum, we are able to further constrain our uncertainties. In general, it is challenging to detect the continua of relatively faint M~stars. Given the proximity of \aumic, we were able to obtain a significant detection of an M~dwarf continuum. Because of this, \aumic\ is an essential benchmark star for understanding UV continua of M~dwarfs, and accurately modeling transmission spectra for planets around these types of stars.

\section{Flare-Affiliated Physical Processes}\label{sec:processes}

Coronal mass ejections (CMEs) are eruptive explosions of magnetized plasma that are ejected from the surface of a star. The typical ejection speeds of the plasma make these events potentially detrimental to planetary atmosphere's chemistry and escape rate \citep{segura10,youngblood17,tilley19,chen21}. \cite{segura10}, \cite{tilley19}, and \cite{chen21} have demonstrated that short-duration bursts of stellar energetic particles (SEPs) from M~dwarf flares can lead to significant O$_3$ depletion on an Earth-like planet without a magnetic field. 

High-energy particles can also compress planetary magnetospheres \citep{cohen14, tilley16}, strip the atmosphere \citep{Lammer07}, and produce harmful atmospheric chemical processes detrimental to surface life. \cite{airapetian20} highlights it is the associated XUV and energetic particles that are heightened and accelerated during CMEs that have the potential to control a planet's climate and habitability. 

However, it is still unclear as to how M~dwarf CMEs differ from solar-type CMEs. \citep{alvaradoGomez16} model magnetic field configurations and CMEs for M~dwarfs. For cases of intermediate and strong strength magnetic fields, it was seen that CMEs can be fully compressed within the magnetic field. This would result in coronal rain rather than being ejected into the local stellar environment. If this is the case for very active M~dwarfs, then CMEs would have potentially negligible effects on short-period exoplanets.

Therefore, understanding the occurrence rate of these processes for \aumic\ is vital for understanding the conditions of the accompanying planets. In this section we describe constraints on coronal mass ejections and nonthermal protons in the stellar atmosphere of for \aumic\ from our COS light curves. 

\subsection{Coronal Mass Ejections Associated with FUV Flares}

Detecting CMEs from spatially unresolved stars is challenging. One promising method to detect CMEs is through the process typically referred to as ``coronal dimming'' \citep{harra16, veronig21, loyd22}. Coronal dimming is caused by the depletion of plasma in the corona of a star as during a CME \citep{hudson96, sterling97}. This effect is observed in EUV wavelengths on the surface of the Sun via spectral tracers of coronal plasma with $T~\sim 10^6$~K \citep{Dissauer18, vanninathan18, mason19}. Recently, \cite{veronig21} searched archival EUV and X-ray observations for CMEs associated with flares on other stars via coronal dimming events. They reported three statistically significant ($\sigma \geq 4.4$) dimming events in X-ray observations of \aumic\ with depths ranging from $12-24$\%. 

Following the methodology outlined in \cite{veronig21}, we searched for dimming events in our light curves created from the \ion{Fe}{12}, \ion{Fe}{19}, and \ion{Fe}{21} emission lines. These lines form at $10^{6.2}$~K, $10^{7.0}$~K, and $10^{7.1}$~K, and trace the quiet and active corona, respectively. We searched for post-flare dimming during Flare~D in these Fe lines. Flare~D is the most energetic flare for which we could reliably establish a quiet pre-flare and post-flare flux level in the white-light data. We define the pre-flare flux in the interval $t_0 - t = 17114-17595$~s, which is 60~s after Flare~C ends. Similarly, we define the post-flare flux in the interval $t_0 - t = 18405-19215$~s, which is 120~s after Flare~D ends. 

We find the pre- and post-flare flux in \ion{Fe}{12} to be $(1.33 \pm 0.58) \times 10^{-13}$ and $(1.31 \pm 0.57) \times 10^{-13}$~erg~s$^{-1}$~cm$^{-2}$. We find the pre- and post-flare flux in \ion{Fe}{19} to be $(4.53 \pm 1.93) \times 10^{-13}$~erg~s$^{-1}$~cm$^{-2}$ and $(6.18 \pm 2.03) \times 10^{-13}$~erg~s$^{-1}$~cm$^{-2}$. We find the pre- and post-flare flux in \ion{Fe}{21} to be $(6.09 \pm 1.25) \times 10^{-13}$~erg~s$^{-1}$~cm$^{-2}$ and $(4.97 \pm 1.08) \times 10^{-13}$~erg~s$^{-1}$~cm$^{-2}$. This indicates that the pre- and post-flare flux for all iron lines searched in this paper are within a $1\sigma$ agreement with each other. Therefore, there is no evidence for coronal dimming associated with Flare D. However, it is not clear whether this non-detection was due to insufficient sensitivity or the lack of a CME. It would be worthwhile to perform a detailed investigation of these two possibilities, but this is beyond the scope of this paper. 

\subsection{Orrall-Zirker Effect}

\cite{orrall76} predicted the existence of additional \lya\ emission during flaring events as a result of nonthermal proton beams. Low-energy ($< 1$\,MeV) protons are challenging to detect because of the lack of affiliated X-ray or microwave radiation. However, it is possible that these protons could interact with and excite chromospheric hydrogen atoms. This process would subsequently result in spontaneous decay and the release of a high-energy photon. The high-energy photons are potentially detectable via flux excess in the red wing of \lya\ \citep{orrall76}. This signature would serve as an observational diagnostic of nonthermal proton beams. 

Observations of \aumic\ with the Goddard High Resolution Spectrograph on HST on 1991 September 3 provided the first statistically significant detection of the Orrall-Zirker Effect on another star. \cite{woodgate92} detected an enhancement in the red wing of \lya\ which lasted approximately 3\,s and contained flux of at least $10^{30}$\,erg\,s$^{-1}$. The excess was seen from $1220 \leq \lambda \leq 1230$\,\angstrom, and is in agreement with the predictions of \cite{orrall76}. Subsequent observations of \aumic\ found no \lya\ enhancement, and placed an upper limit on the energy of the beam to $\leq 10^{29}$\,erg\,s$^{-1}$ \citep{robinson93}.

To determine if there was an affiliated proton beam in our observations, we searched for enhancement in the blue and red wings of \lya, respectively. Specifically, we followed the prescription for the observational requirements of a true event presented in Section~2 of \cite{woodgate92}. We created 1~second light curves from the third orbit of Visit~1 to search for an affiliated proton beam around both discrete peaks of Flare~B from $1202-1204$\,\angstrom\ and $1222-1227$\,\angstrom. Although we discovered a stronger count enhancement in the blue-wing than the red-wing, the overall detection of enhancement was non-significant. Therefore, we conclude that there is no evidence for the Orrall-Zirker Effect in the observations presented in this paper. Future HST-STIS observations of Ly$\alpha$ would be a promising avenue to observe this process in bright stars like \aumic.

\section{Conclusions} \label{sec:conclusion}

In this paper, we presented HST/COS observations of \nflares\ flares on \aumic. Our observations spanned ten orbits over two visits. We summarize our main takeaways below. 

\begin{enumerate}
 \item In Section~\ref{sec:analysis}, we measured flare energies ranging from $1 - 24 \times 10^{30}$~erg. The latter are comparable to the Sun's Carrington event \citep{carrington1859}. We discovered a UV flare rate of $\sim 2$\,flares/hour, which is significantly greater than the one presented in \cite{gilbert22}. This discrepancy is likely due to the difference in bandpasses between HST/COS and TESS. Our findings suggest that the FUV flare rate of low-mass stars is higher than in the optical/IR. This is because lower-energy flares are easier to observe in the UV than optical due to decreased photospheric background level. 
 
 \item In Section~\ref{subsec:speclc}, we created spectroscopic light curves for a range of atmospheric formation temperatures, and found that all flares have the strongest measured energies at log$_{10} (T_\textrm{form} [$K$]) = 4.8$. We also found a ubiquitous and persistent redshift in the line profiles, which could be due to chromospheric condensation \citep{hawley03}.
 
 \item In Section~\ref{subsec:continua}, we estimated a blackbody continuum temperature of $\sim 15,000$\,K at $\lambda > 1100$\,\angstrom\ . However, it is important to note that the quiescent blackbody temperature is comparable or greater than those measured during Flares~B, D, J, K, and M. Therefore, it is not clear that a blackbody is the best model to fit to the flare continuum.
 
 \item Additionally in Section~\ref{subsec:continua}, we identified a steep increase in continuum flux during the observed flares at $\lambda < 1100$\,\angstrom. This was best-fit with a thermal bremsstrahlung profile with $9 \leq \textrm{log}_{10}(T) \leq 11$, similar to the measurements presented in \cite{redfield02}. If this interpretation proves correct, there would be an enhancement of EUV flux from \aumic. This enhancement could produce an increase in the atmospheric mass loss of \planetboth. 
 
 \item In Section~\ref{sec:panchromatic}, we created a full panchromatic spectrum of \aumic\ with archival XMM-Newton, FUSE, IUE, and HARPS-N observations. For wavelengths that have not been observed ($40 \geq \lambda \leq 900$\,\angstrom\ ), we fit a differential emission measurement (DEM) model to the COS observations presented in this paper. Similarly, We filled in redward of the HARPS-N observations with a PHOENIX stellar model. This SED will be available for use in atmospheric modeling at \url{https://archive.stsci.edu/prepds/muscles/}.
 
 \item In Section~\ref{subsubsec:massloss}, we calculated an approximate atmospheric mass-loss rate due to photoevaporation for \planetb. In the calculation, we used the estimated high-energy luminosity from our panchromatic spectrum and injected flares with energies ranging from $E = 10^{29-34}$\,erg. We found that flares could temporarily increase the mass-loss for \planetb\ to five orders of magnitude above the current non-detection limits of $10^8$\,g\,s$^{-1}$ \citep{hirano20}.
 
 \item In Section~\ref{subsubsec:obssig}, we modeled the optical through mid-IR transmission spectra for \planetboth\ using our newly created panchromatic spectrum. Additionally, we modeled the temperature-pressure profiles for both planets in quiescence. We estimate transit depth differences between the equilibrium and disequilibrium models of $\approx 70-200$\,ppm, depending on the wavelength range. These differences could be observable with two transits per planet using JWST.
 
\end{enumerate}

\subsection{Future Work}


The photoevaporative mass-loss estimated in this work both including and excluding flare events suggests \planetboth\ could lose 30-50\% of their present-day atmospheres. Compared to larger, younger planets \citep[e.g.][]{david19_v1298all, rizzuto20}, the planets orbiting \aumic\ will not undergo significant radial evolution as the system ages. Atmospheric escape of neutral hydrogen may be observed via the identification of absorption in the wings of \lya\ (e.g., \citealt{bourrier18}). It is unlikely that flare-driven atmospheric mass loss would be observed due to geometric constraints on the location of the flare with respect to the orbital plane. And, simulations have shown that the radiation from flares alone produce only minor ($<0.1\%$) differences in transit shapes \citep{hazra22}. 

However, it is possible in the presence of a flare and affiliated CME that these processes will be detectable for \aumic\ because it is close and bright. Future transit observations of \planetboth\ would have the required signal-to-noise to confidently detect increases in \lya\ depths. This would correspond to mass loss rates of $\dot{M} \geq 10^{10}$\,g\,s$^{-1}$ (\rockcliffep). Unfortunately, contamination in \lya\ from the interstellar medium makes constraining the atmospheric composition of these planets challenging. Simultaneous observations of other dominant atmospheric species such as the metastable \ion{He}{1} triplet feature at 10830\,\angstrom, or the strong \ion{O}{1} and \ion{C}{2} lines in the FUV would provide additional constraints on the atmospheric composition. 

The effects of flares and XUV irradiation have also been considered for old ($t_\textrm{age} > 1$~Gyr) stars. The contribution of flares to the high-energy radiation has been shown to remove $\sim$90 Earth-atmospheres within a Gyr for old and inactive M dwarfs \citep{france20}. The full extent of flare-driven atmospheric removal for more massive planets ($M > 10 M_\oplus$) with H$_2$-rich envelopes such as Jupiter and Neptune has not been fully investigated. Modeling the differences in observed transmission spectra during flares of varying energies would help interpret upcoming JWST observations of \planetboth.

Understanding the contribution of thermal and non-thermal processes to atmospheric mass loss is unknown for exoplanets. The detection of radio emission from \planetboth\ would yield constraints on the planetary magnetic field strengths and, in turn, how well shielded the planets are from strong stellar winds and CMEs. While there is weak evidence of radio emission from hot Jupiters \citep[e.g.][]{lecavelier13, deGasperin20, narang21}, \aumic\ has potential. \cite{kavanagh21} simulated Alfv\'{e}n wave-driven stellar wind models to investigate potential auroral signals in the stellar corona from interactions with \planetboth. In the low mass-loss rate ($\sim 27 \dot{M_\odot}$) scenario, \planetboth\ are orbiting sub-Alfv\'{e}nifcally, and \planetb\ could produce time-varying radio emission from $\sim 10$\,Mz $- 3$\,GHz at detectable levels.

The strong FUV continuum increase at $\lambda \leq 1100$\,\angstrom\ is readily seen during flares in our HST/COS observations (Figure~\ref{fig:bbs}). We tentatively attribute this observational feature to thermal bremsstrahlung emission. Flare observations in the COS FUV modes covering even shorter wavelengths would help constrain the overall contribution of thermal bremsstrahlung to the flare energy output. However, the high-energy luminosity calculated from our DEM model and used as an input to our atmospheric mass-loss calculation does not account for this additional emission. Accurate treatment of this additional emission from thermal bremsstrahlung may result in more stringent constraints on the contribution of flare energies to young atmospheric removal.

\subsection{Software \& Data Availability}

We have packaged all analysis tools that were used throughout this work on our GitHub repository: \href{https://github.com/afeinstein20/cos_flares/tree/paper-version}{@afeinstein20/cos\_flares/tree/paper-version}. Under this repository, we provide demonstration Jupyter notebooks for setting up the code for future HST/COS flare observations. Notebooks for specific figures are highlighted with the \codeicon\ icon throughout the paper. 

We provide a variety of data products from this analysis, which has been packaged on Zenodo DOI:10.5281/zenodo.6386814.\footnote{\url{https://doi.org/10.5281/zenodo.6386814
}} We provide a machine-readable version of Table~\ref{tab:linelist}, our DEM model for \aumic\ in quiescence, all spectra used to create the SED for AU Mic (Figure~\ref{fig:panchromatic}) and our modeled transmission spectra for \planetboth\ (Figure~\ref{fig:transmission}). We provide documentation for using these data products on Zenodo as well.

\vspace{5mm}

We thank Thomas Berger, Alex Brown, Patrick Behr, Megan Mansfield, and Yamila Miguel for insightful conversations. We thank the anonymous reviewer for their insightful and thorough comments, which have improved the quality of this manuscript.

Support for Program 16164 was provided by NASA through a grant from the Space Telescope Science Institute, which is operated by the Association of Universities for Research in Astronomy, Incorporated, under NASA contract NAS5-26555.

ADF acknowledges support by the National Science Foundation Graduate Research Fellowship Program under Grant No. (DGE-1746045). DJT and EMRK acknowledge funding from the NSF AAG program (grant \#2009095).

This research has made use of NASA's Astrophysics Data System Bibliographic Services. CHIANTI is a collaborative project involving George Mason University, the University of Michigan (USA), University of Cambridge (UK) and NASA Goddard Space Flight Center (USA). This research made use of PlasmaPy version 0.6.0, a community-developed open source Python package for plasma science \citep{plasmapy}.


\appendix
\restartappendixnumbering

\section{Supplemental Material}\label{appendix:lightcurves}

\subsection{Full Spectroscopic Light Curves}\label{appendix:lcs}
\begin{figure*}[!ht]
\begin{center}
\includegraphics[width=0.9\textwidth,trim={0.25cm 0 0 0}]{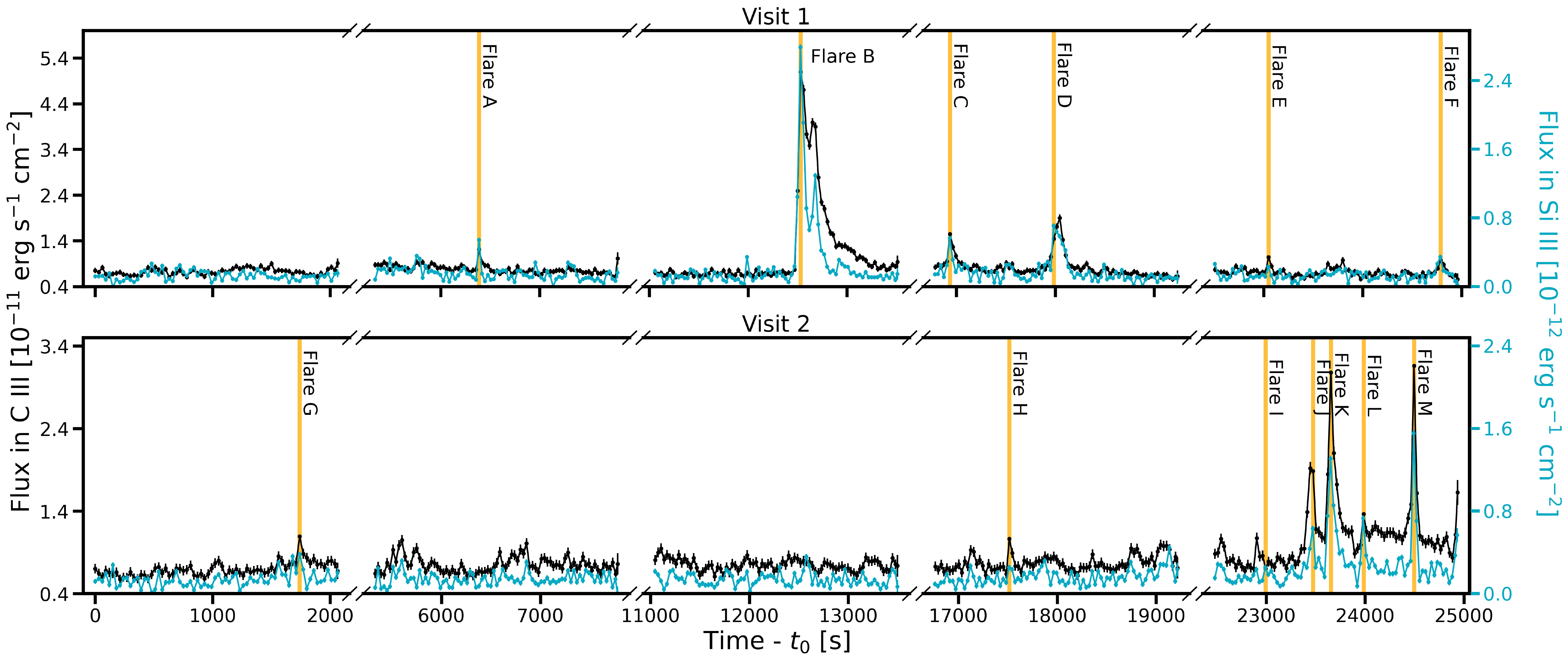}
\caption{Light curves of the \ion{C}{3} emission line at 1175.95\,\angstrom\ (black) and the \ion{Si}{3} emission line at 1294.55\,\angstrom\ (blue) used to identify flares in the data. Flares identified are labeled with vertical orange lines. \label{fig:id_lc}}
\end{center}
\end{figure*}

We used the \ion{C}{3} at $\lambda = 1175.95$\,\angstrom\ and \ion{Si}{3} $\lambda = 1294.55$\,\angstrom\ spectroscopic light curves to identify flares in our data set (Figure~\ref{fig:id_lc}). These are the same methods presented in \cite{woodgate92}.

\subsection{Quiescent Spectrum}

\begin{figure*}[p]
\begin{center}
\includegraphics[width=\textwidth,trim={0.25cm 0 0 0}]{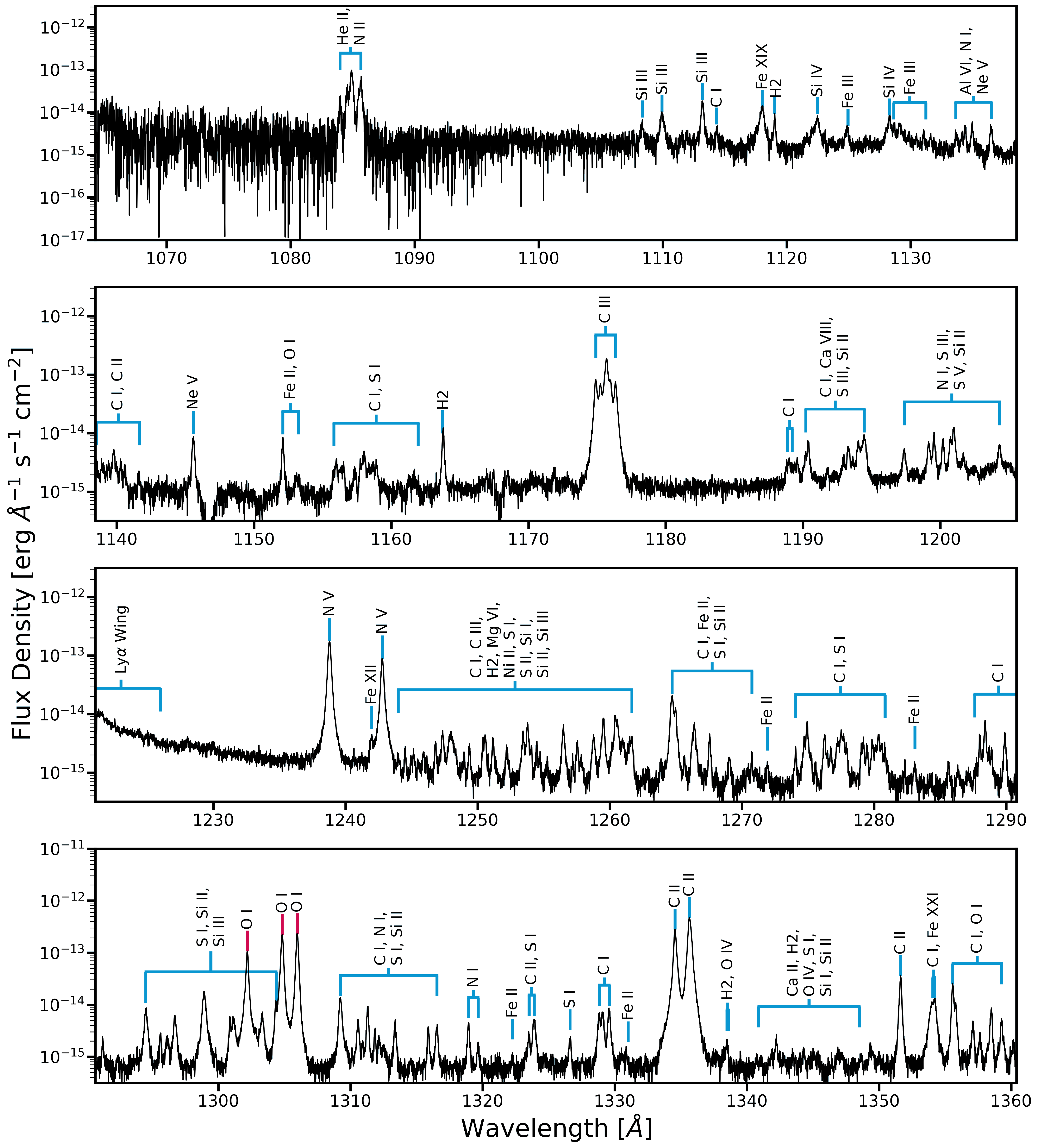}
\caption{The average quiescent spectrum for \aumic. We removed time intervals that fall within the highlighted yellow regions in Figure~\ref{fig:lightcurve}. We labeled all known emission lines seen in our spectrum. Emission features marked with pink lines (\ion{O}{1} triplet) are partially contaminated by air glow. We present all measured line centers and flux values for these emission features in Table~\ref{tab:lines}. \href{https://github.com/afeinstein20/cos_flares/blob/paper-version/notebooks/labeled_quiescent_spectrum.ipynb}{\codeicon} \label{fig:spectrum}}
\end{center}
\end{figure*}

We present the entire mean quiescent spectrum for \aumic, with labeled identified emission features, in Figure~\ref{fig:spectrum}. We grouped dense regions of lines together in the Figure, and provide measured flux values for all identified lines in Table~\ref{tab:linelist}. Table~\ref{tab:linelist}, which contains all identified lines with the following parameters: the measured rest wavelength ($\lambda_\textrm{rest}$), the observed wavelength ($\lambda_\textrm{obs}$), the velocity shift between rest and observed wavelengths in [km s$^{-1}$], the measured flux in quiescence and during Flare~B in $10^{-15}$erg\,s$^{-1}$\,cm$^{-2}$, and the full-width half-maximum (FWHM)[km s$^{-1}$] of the line in quiescence and during Flare~B.

\subsection{Continuum Regions}\label{appendix:continuum}

We identified the continuum regions of the spectra by-eye, and defined these regions to have no emission features. These are the following wavelength regions we define as the continuum: [1067.506, 1070.062], [1074.662, 1076.533], [1078.881, 1082.167], [1087.828, 1090.035], [1103.787, 1107.862], [1110.500, 1112.946], [1113.618, 1117.377], [1119.548, 1121.622], [1125.255, 1126.923], [1140.873, 1145.141], [1146.285, 1151.544], [1152.602, 1155.579], [1159.276, 1163.222], [1164.565, 1173.959], [1178.669, 1188.363], [1195.162, 1196.864], [1201.748, 1203.862], [1227.056, 1236.921], [1262.399, 1263.967], [1268.559, 1273.974], [1281.396, 1287.493], [1290.494, 1293.803], [1307.064, 1308.703], [1319.494, 1322.910], [1330.349, 1332.884], [1337.703, 1341.813], and [1341.116, 1350.847] \angstrom.

\newpage
\startlongtable
\begin{deluxetable}{l r r r r r r r}
\tabletypesize{\footnotesize}
\tablecaption{Complete line list for emission features present in our \textit{Hubble}/COS spectra for AU~Mic.\label{tab:lines}}
\tablehead{
\colhead{Ion} & \colhead{$\lambda_\textrm{rest}$} & \colhead{$\lambda_\textrm{obs}$} & \colhead{Velocity Shift} & \colhead{Flux (Quiescent)} & \colhead{FWHM (Quiescent)} & \colhead{Flux (Flare B)} & \colhead{FWHM (Flare B)}\\
\colhead{} & \colhead{[\AA]} & \colhead{[\AA]} & \colhead{[km s$^{-1}$]} & \colhead{[10$^{-15}$ erg s$^{-1}$ cm$^{-2}$]} & \colhead{[km s$^{-1}$]} & \colhead{[10$^{-15}$ erg s$^{-1}$ cm$^{-2}$]} & \colhead{[km s$^{-1}$]}}
\startdata
N II & 1083.99 & 1083.99 & -2.67 & 4.26 $\pm$ 5.4 & 0.26 $\pm$ 0.02 & 5.82 $\pm$ 21.47 & 0.17 $\pm$ 0.03 \\
N II * & 1084.58 & 1084.57 & -2.67 & 8.22 $\pm$ 4.54 & 0.29 $\pm$ 0.02 & 10.96 $\pm$ 17.9 & 0.33 $\pm$ 0.09 \\
N II & 1085.54 & 1085.56 & 5.33 & 16.98 $\pm$ 5.17 & 0.32 $\pm$ 0.01 & 21.37 $\pm$ 20.68 & 0.36 $\pm$ 0.04 \\
N II & 1085.71 & 1085.66 & -13.32 & 16.98 $\pm$ 5.17 & 0.32 $\pm$ 0.01 & 21.37 $\pm$ 20.68 & 0.36 $\pm$ 0.04 \\
Si III * & 1108.36 & 1108.35 & -2.61 & 2.22 $\pm$ 0.82 & 0.51 $\pm$ 0.03 & 4.81 $\pm$ 3.26 & 0.45 $\pm$ 0.04 \\
Si III * & 1109.94 & 1109.95 & 0.0 & 3.69 $\pm$ 0.75 & 0.45 $\pm$ 0.01 & 7.29 $\pm$ 3.0 & 0.36 $\pm$ 0.02 \\
Si III * & 1113.2 & 1113.21 & 0.0 & 4.92 $\pm$ 0.63 & 0.3 $\pm$ 0.01 & 11.27 $\pm$ 2.58 & 0.46 $\pm$ 0.03 \\
C I & 1114.39 & 1114.35 & -10.38 & 1.64 $\pm$ 0.59 & 0.62 $\pm$ 0.05 & 2.09 $\pm$ 2.37 & 0.49 $\pm$ 0.07 \\
Fe XIX * & 1118.06 & 1118.02 & -12.94 & 5.43 $\pm$ 0.52 & 0.45 $\pm$ 0.01 & 5.34 $\pm$ 2.07 & 0.5 $\pm$ 0.04 \\
Si IV * & 1122.49 & 1122.47 & -5.15 & 3.29 $\pm$ 0.45 & 0.57 $\pm$ 0.02 & 6.19 $\pm$ 1.82 & 0.71 $\pm$ 0.07 \\
Fe III & 1124.88 & 1124.94 & 18.0 & 2.06 $\pm$ 0.44 & 0.6 $\pm$ 0.03 & 3.37 $\pm$ 1.75 & 0.58 $\pm$ 0.06 \\
Si IV * & 1128.34 & 1128.3 & -12.82 & 3.06 $\pm$ 0.36 & 0.54 $\pm$ 0.03 & 6.21 $\pm$ 1.45 & 0.5 $\pm$ 0.05 \\
Al VI & 1133.68 & 1133.62 & -17.86 & 1.26 $\pm$ 0.31 & 0.49 $\pm$ 0.03 & 1.54 $\pm$ 1.23 & 1.68 $\pm$ 2.35 \\
N I & 1134.16 & 1134.14 & -5.1 & 0.73 $\pm$ 0.21 & 0.3 $\pm$ 0.02 & 0.9 $\pm$ 0.86 & 0.77 $\pm$ 1.18 \\
N I & 1134.4 & 1134.39 & -2.55 & 0.91 $\pm$ 0.21 & 0.22 $\pm$ 0.01 & 1.07 $\pm$ 0.86 & 0.33 $\pm$ 0.09 \\
N I & 1134.98 & 1134.95 & -7.65 & 1.52 $\pm$ 0.3 & 0.37 $\pm$ 0.02 & 2.02 $\pm$ 1.2 & 0.29 $\pm$ 0.03 \\
Ne V * & 1136.49 & 1136.49 & 0.0 & 1.38 $\pm$ 0.29 & 0.36 $\pm$ 0.01 & 1.62 $\pm$ 1.16 & 0.7 $\pm$ 0.17 \\
C I & 1138.95 & 1138.95 & 0.0 & 0.97 $\pm$ 0.23 & 0.46 $\pm$ 0.03 & 1.31 $\pm$ 0.92 & 0.47 $\pm$ 0.13 \\
C I & 1139.81 & 1139.78 & -7.61 & 1.61 $\pm$ 0.25 & 0.42 $\pm$ 0.02 & 2.44 $\pm$ 0.99 & 0.38 $\pm$ 0.04 \\
C I & 1140.35 & 1140.26 & -22.83 & 0.96 $\pm$ 0.23 & 0.48 $\pm$ 0.05 & 1.06 $\pm$ 0.93 & 0.38 $\pm$ 0.07 \\
C I & 1140.62 & 1140.6 & -5.07 & 0.8 $\pm$ 0.23 & 0.43 $\pm$ 0.03 & 0.88 $\pm$ 0.93 & 0.33 $\pm$ 0.07 \\
C II * & 1141.68 & 1141.64 & -10.14 & 0.63 $\pm$ 0.23 & 0.46 $\pm$ 0.04 & 0.78 $\pm$ 0.91 & 1.28 $\pm$ 2.11 \\
Ne V * & 1145.58 & 1145.57 & -2.53 & 1.95 $\pm$ 0.22 & 0.24 $\pm$ 0.01 & 2.15 $\pm$ 0.88 & 0.27 $\pm$ 0.02 \\
C I & 1157.4 & 1157.35 & -12.5 & 0.93 $\pm$ 0.22 & 0.52 $\pm$ 0.03 & 1.41 $\pm$ 0.88 & 0.49 $\pm$ 0.06 \\
C III & 1174.88 & 1174.9 & 4.92 & 24.01 $\pm$ 0.31 & 0.29 $\pm$ 0.01 & 66.86 $\pm$ 1.45 & 0.45 $\pm$ 0.02 \\
C III & 1175.24 & 1175.24 & 0.0 & 17.09 $\pm$ 0.19 & 0.35 $\pm$ 0.02 & 43.11 $\pm$ 0.95 & 0.52 $\pm$ 0.04 \\
C III & 1176.37 & 1176.34 & -7.38 & 21.45 $\pm$ 0.31 & 0.31 $\pm$ 0.01 & 56.99 $\pm$ 1.38 & 0.57 $\pm$ 0.02 \\
S III * & 1190.17 & 1190.2 & 7.29 & 1.25 $\pm$ 0.14 & 0.57 $\pm$ 0.06 & 2.35 $\pm$ 0.56 & 0.54 $\pm$ 0.13 \\
C I & 1191.84 & 1191.79 & -14.56 & 0.99 $\pm$ 0.16 & 0.67 $\pm$ 0.06 & 1.67 $\pm$ 0.63 & 0.46 $\pm$ 0.06 \\
C I & 1193.0 & 1192.99 & -2.42 & 1.25 $\pm$ 0.14 & 0.51 $\pm$ 0.04 & 2.0 $\pm$ 0.57 & 0.48 $\pm$ 0.08 \\
C I & 1193.29 & 1193.3 & 2.42 & 1.51 $\pm$ 0.13 & 0.32 $\pm$ 0.01 & 2.44 $\pm$ 0.52 & 1.0 $\pm$ 0.99 \\
C I & 1193.68 & 1193.61 & -16.96 & 0.86 $\pm$ 0.11 & 0.36 $\pm$ 0.03 & 1.66 $\pm$ 0.46 & 0.54 $\pm$ 0.29 \\
Ca VIII * & 1194.04 & 1194.04 & 0.0 & 2.16 $\pm$ 0.14 & 0.43 $\pm$ 0.02 & 3.93 $\pm$ 0.58 & 0.4 $\pm$ 0.03 \\
Si II * & 1194.45 & 1194.46 & 2.42 & 2.96 $\pm$ 0.16 & 0.42 $\pm$ 0.01 & 5.72 $\pm$ 0.65 & 0.44 $\pm$ 0.04 \\
Si II * & 1197.4 & 1197.37 & -7.25 & 1.96 $\pm$ 0.17 & 0.46 $\pm$ 0.02 & 3.23 $\pm$ 0.67 & 0.53 $\pm$ 0.05 \\
S V * & 1199.2 & 1199.16 & -9.65 & 2.26 $\pm$ 0.16 & 0.44 $\pm$ 0.02 & 3.23 $\pm$ 0.62 & 0.51 $\pm$ 0.04 \\
N I & 1199.55 & 1199.55 & -2.41 & 2.3 $\pm$ 0.14 & 0.29 $\pm$ 0.01 & 3.4 $\pm$ 0.57 & 0.35 $\pm$ 0.03 \\
N I & 1200.22 & 1200.2 & -4.82 & 1.98 $\pm$ 0.14 & 0.29 $\pm$ 0.01 & 3.09 $\pm$ 0.56 & 0.48 $\pm$ 0.06 \\
N I & 1200.71 & 1200.75 & 9.64 & 2.05 $\pm$ 0.12 & 0.33 $\pm$ 0.01 & 3.37 $\pm$ 0.51 & 0.58 $\pm$ 0.16 \\
S III & 1200.99 & 1200.99 & -2.41 & 3.11 $\pm$ 0.14 & 0.34 $\pm$ 0.02 & 6.44 $\pm$ 0.59 & 0.44 $\pm$ 0.04 \\
S III * & 1201.73 & 1201.69 & -12.04 & 1.95 $\pm$ 0.16 & 0.77 $\pm$ 0.05 & 3.95 $\pm$ 0.66 & 0.73 $\pm$ 0.1 \\
S V * & 1204.3 & 1204.31 & 2.4 & 2.49 $\pm$ 0.16 & 0.53 $\pm$ 0.02 & 6.2 $\pm$ 0.67 & 0.72 $\pm$ 0.06 \\
N V * & 1238.82 & 1238.79 & -7.01 & 53.45 $\pm$ 0.35 & 0.26 $\pm$ 0.0 & 85.24 $\pm$ 1.51 & 0.35 $\pm$ 0.01 \\
Fe XII * & 1241.95 & 1241.98 & 6.99 & 1.69 $\pm$ 0.16 & 0.56 $\pm$ 0.03 & 3.16 $\pm$ 0.65 & 0.8 $\pm$ 0.15 \\
N V * & 1242.8 & 1242.79 & -4.66 & 27.19 $\pm$ 0.3 & 0.26 $\pm$ 0.0 & 41.19 $\pm$ 1.23 & 0.32 $\pm$ 0.01 \\
C I & 1244.51 & 1244.51 & 0.0 & 0.58 $\pm$ 0.13 & 0.31 $\pm$ 0.02 & 0.97 $\pm$ 0.52 & 0.35 $\pm$ 0.05 \\
C I & 1246.87 & 1246.81 & -16.24 & 0.82 $\pm$ 0.14 & 0.32 $\pm$ 0.02 & 1.35 $\pm$ 0.57 & 0.4 $\pm$ 0.05 \\
C III & 1247.41 & 1247.35 & -16.23 & 1.3 $\pm$ 0.14 & 0.32 $\pm$ 0.01 & 3.42 $\pm$ 0.59 & 0.32 $\pm$ 0.02 \\
C I & 1247.86 & 1247.94 & 18.55 & 2.71 $\pm$ 0.21 & 0.68 $\pm$ 0.02 & 3.85 $\pm$ 0.83 & 0.99 $\pm$ 0.11 \\
C I & 1248.0 & 1247.99 & -2.32 & 2.7 $\pm$ 0.21 & 0.71 $\pm$ 0.02 & 3.72 $\pm$ 0.83 & 1.16 $\pm$ 0.19 \\
C I & 1249.41 & 1249.37 & -9.26 & 0.84 $\pm$ 0.16 & 0.36 $\pm$ 0.02 & 1.37 $\pm$ 0.64 & 0.41 $\pm$ 0.05 \\
S II * & 1250.58 & 1250.56 & -4.63 & 1.51 $\pm$ 0.16 & 0.41 $\pm$ 0.02 & 2.69 $\pm$ 0.65 & 0.43 $\pm$ 0.03 \\
Si II * & 1251.16 & 1251.16 & 0.0 & 1.04 $\pm$ 0.15 & 0.33 $\pm$ 0.02 & 1.78 $\pm$ 0.6 & 0.33 $\pm$ 0.02 \\
C I & 1252.21 & 1252.21 & -2.31 & 0.86 $\pm$ 0.15 & 0.39 $\pm$ 0.02 & 1.13 $\pm$ 0.6 & 0.31 $\pm$ 0.03 \\
C I & 1253.47 & 1253.44 & -6.92 & 1.3 $\pm$ 0.15 & 0.36 $\pm$ 0.02 & 1.84 $\pm$ 0.58 & 0.37 $\pm$ 0.04 \\
S II * & 1253.8 & 1253.78 & -2.31 & 1.5 $\pm$ 0.13 & 0.26 $\pm$ 0.01 & 2.28 $\pm$ 0.51 & 0.3 $\pm$ 0.02 \\
C I & 1254.51 & 1254.48 & -6.92 & 0.7 $\pm$ 0.13 & 0.29 $\pm$ 0.02 & 1.01 $\pm$ 0.5 & 0.31 $\pm$ 0.03 \\
Si I & 1255.28 & 1255.27 & -4.61 & 0.51 $\pm$ 0.14 & 0.46 $\pm$ 0.03 & 0.61 $\pm$ 0.57 & 1.02 $\pm$ 0.88 \\
Mg VI * & 1256.37 & 1256.48 & 25.33 & 1.83 $\pm$ 0.19 & 0.32 $\pm$ 0.01 & 2.51 $\pm$ 0.74 & 0.4 $\pm$ 0.02 \\
C I & 1256.5 & 1256.49 & 0.0 & 0.7 $\pm$ 0.12 & 0.24 $\pm$ 0.01 & 1.16 $\pm$ 0.47 & 0.43 $\pm$ 0.14 \\
Si I & 1258.78 & 1258.78 & 0.0 & 1.39 $\pm$ 0.17 & 0.38 $\pm$ 0.01 & 2.34 $\pm$ 0.66 & 0.42 $\pm$ 0.03 \\
S II * & 1259.53 & 1259.52 & -2.3 & 2.62 $\pm$ 0.19 & 0.34 $\pm$ 0.01 & 4.62 $\pm$ 0.76 & 0.48 $\pm$ 0.02 \\
Si II * & 1260.44 & 1260.43 & 0.0 & 3.86 $\pm$ 0.19 & 0.48 $\pm$ 0.02 & 6.25 $\pm$ 0.77 & 0.55 $\pm$ 0.03 \\
C I & 1261.72 & 1261.67 & -11.46 & 1.81 $\pm$ 0.19 & 0.54 $\pm$ 0.02 & 2.84 $\pm$ 0.74 & 0.61 $\pm$ 0.05 \\
Si II * & 1264.74 & 1264.72 & -4.57 & 8.82 $\pm$ 0.26 & 0.48 $\pm$ 0.01 & 14.0 $\pm$ 1.05 & 0.6 $\pm$ 0.02 \\
Si II & 1265.0 & 1264.97 & -6.86 & 8.56 $\pm$ 0.24 & 0.48 $\pm$ 0.01 & 13.36 $\pm$ 0.98 & 0.58 $\pm$ 0.02 \\
C I & 1267.6 & 1267.56 & -9.13 & 0.95 $\pm$ 0.15 & 0.24 $\pm$ 0.01 & 1.26 $\pm$ 0.59 & 0.29 $\pm$ 0.02 \\
S I & 1269.06 & 1269.03 & -6.84 & 0.65 $\pm$ 0.16 & 0.46 $\pm$ 0.02 & 0.75 $\pm$ 0.62 & 0.44 $\pm$ 0.06 \\
S I & 1270.78 & 1270.75 & -6.83 & 0.5 $\pm$ 0.13 & 0.31 $\pm$ 0.02 & 0.76 $\pm$ 0.52 & 0.4 $\pm$ 0.06 \\
Fe II & 1271.98 & 1271.92 & -13.65 & 0.47 $\pm$ 0.14 & 0.45 $\pm$ 0.03 & 0.64 $\pm$ 0.57 & 0.48 $\pm$ 0.1 \\
C I & 1274.11 & 1274.07 & -11.35 & 0.63 $\pm$ 0.14 & 0.34 $\pm$ 0.02 & 0.96 $\pm$ 0.57 & 0.3 $\pm$ 0.03 \\
C I & 1274.98 & 1274.93 & -11.34 & 1.59 $\pm$ 0.13 & 0.25 $\pm$ 0.01 & 2.19 $\pm$ 0.54 & 0.2 $\pm$ 0.02 \\
C I & 1276.29 & 1276.26 & -9.07 & 1.25 $\pm$ 0.15 & 0.36 $\pm$ 0.02 & 2.03 $\pm$ 0.6 & 0.35 $\pm$ 0.03 \\
C I & 1276.75 & 1276.73 & -4.53 & 0.77 $\pm$ 0.14 & 0.43 $\pm$ 0.03 & 1.36 $\pm$ 0.56 & 0.51 $\pm$ 0.08 \\
C I & 1279.5 & 1279.46 & -9.04 & 0.62 $\pm$ 0.12 & 0.28 $\pm$ 0.02 & 1.05 $\pm$ 0.47 & 0.28 $\pm$ 0.04 \\
C I & 1279.89 & 1279.88 & -2.26 & 0.67 $\pm$ 0.11 & 0.26 $\pm$ 0.02 & 1.24 $\pm$ 0.46 & 0.26 $\pm$ 0.04 \\
C I & 1280.33 & 1280.35 & 4.52 & 1.08 $\pm$ 0.12 & 0.33 $\pm$ 0.02 & 1.64 $\pm$ 0.48 & 0.3 $\pm$ 0.03 \\
C I & 1280.85 & 1280.82 & -6.78 & 0.63 $\pm$ 0.12 & 0.3 $\pm$ 0.02 & 1.12 $\pm$ 0.48 & 0.4 $\pm$ 0.09 \\
Fe II & 1283.06 & 1283.09 & 4.51 & 0.43 $\pm$ 0.14 & 0.42 $\pm$ 0.02 & 0.87 $\pm$ 0.56 & 0.42 $\pm$ 0.07 \\
C I & 1288.04 & 1288.01 & -6.74 & 1.0 $\pm$ 0.14 & 0.31 $\pm$ 0.02 & 1.4 $\pm$ 0.55 & 0.31 $\pm$ 0.03 \\
C I & 1288.42 & 1288.41 & -4.49 & 1.56 $\pm$ 0.14 & 0.23 $\pm$ 0.01 & 2.34 $\pm$ 0.56 & 0.22 $\pm$ 0.01 \\
C I & 1288.71 & 1288.7 & -2.24 & 0.43 $\pm$ 0.1 & 0.23 $\pm$ 0.02 & 0.69 $\pm$ 0.39 & 0.47 $\pm$ 0.4 \\
C I & 1288.92 & 1288.89 & -6.73 & 0.43 $\pm$ 0.11 & 0.22 $\pm$ 0.02 & 0.7 $\pm$ 0.42 & 0.26 $\pm$ 0.05 \\
C I & 1289.98 & 1289.9 & -15.7 & 1.18 $\pm$ 0.18 & 0.25 $\pm$ 0.01 & 1.68 $\pm$ 0.7 & 0.29 $\pm$ 0.02 \\
C I & 1291.3 & 1291.26 & -8.96 & 0.63 $\pm$ 0.16 & 0.45 $\pm$ 0.03 & 0.72 $\pm$ 0.64 & 0.53 $\pm$ 0.1 \\
Si III & 1294.58 & 1294.5 & -17.88 & 2.82 $\pm$ 0.21 & 0.34 $\pm$ 0.01 & 7.14 $\pm$ 0.86 & 0.44 $\pm$ 0.02 \\
S I & 1295.65 & 1295.61 & -8.93 & 0.8 $\pm$ 0.17 & 0.4 $\pm$ 0.03 & 1.65 $\pm$ 0.67 & 0.53 $\pm$ 0.07 \\
S I & 1296.16 & 1296.1 & -13.39 & 0.86 $\pm$ 0.16 & 0.47 $\pm$ 0.02 & 1.89 $\pm$ 0.65 & 0.79 $\pm$ 0.2 \\
Si III & 1296.77 & 1296.69 & -20.08 & 1.81 $\pm$ 0.17 & 0.33 $\pm$ 0.01 & 5.07 $\pm$ 0.69 & 0.49 $\pm$ 0.03 \\
Si III & 1298.96 & 1298.93 & -8.91 & 6.83 $\pm$ 0.28 & 0.34 $\pm$ 0.01 & 16.83 $\pm$ 1.16 & 0.5 $\pm$ 0.01 \\
S I & 1300.91 & 1300.9 & -2.22 & 2.99 $\pm$ 0.22 & 0.63 $\pm$ 0.02 & 6.55 $\pm$ 0.89 & 0.74 $\pm$ 0.04 \\
Si III & 1303.32 & 1303.31 & -2.22 & 2.44 $\pm$ 0.19 & 0.44 $\pm$ 0.01 & 6.66 $\pm$ 0.78 & 0.64 $\pm$ 0.04 \\
Si II * & 1304.37 & 1304.37 & 0.0 & 2.62 $\pm$ 0.16 & 0.26 $\pm$ 0.01 & 4.76 $\pm$ 0.64 & 0.45 $\pm$ 0.05 \\
Si II * & 1309.28 & 1309.23 & -11.05 & 4.48 $\pm$ 0.24 & 0.28 $\pm$ 0.01 & 7.38 $\pm$ 0.97 & 0.49 $\pm$ 0.02 \\
N I & 1310.54 & 1310.57 & 6.62 & 1.26 $\pm$ 0.16 & 0.28 $\pm$ 0.01 & 2.07 $\pm$ 0.62 & 0.29 $\pm$ 0.02 \\
C I & 1310.64 & 1310.56 & -17.66 & 1.77 $\pm$ 0.15 & 0.19 $\pm$ 0.01 & 2.76 $\pm$ 0.62 & 0.22 $\pm$ 0.01 \\
N I & 1310.94 & 1310.9 & -8.83 & 0.5 $\pm$ 0.13 & 0.45 $\pm$ 0.06 & 0.92 $\pm$ 0.5 & 0.7 $\pm$ 0.53 \\
C I & 1311.36 & 1311.29 & -17.65 & 1.77 $\pm$ 0.15 & 0.19 $\pm$ 0.01 & 2.76 $\pm$ 0.62 & 0.22 $\pm$ 0.01 \\
C I & 1311.92 & 1311.85 & -15.44 & 0.73 $\pm$ 0.14 & 0.24 $\pm$ 0.01 & 1.07 $\pm$ 0.54 & 0.23 $\pm$ 0.02 \\
C I & 1313.39 & 1313.38 & -2.2 & 1.28 $\pm$ 0.19 & 0.27 $\pm$ 0.01 & 1.73 $\pm$ 0.75 & 0.32 $\pm$ 0.03 \\
C I & 1315.88 & 1315.89 & 2.2 & 0.8 $\pm$ 0.15 & 0.23 $\pm$ 0.01 & 1.11 $\pm$ 0.61 & 0.26 $\pm$ 0.03 \\
S I & 1316.54 & 1316.54 & -2.2 & 1.07 $\pm$ 0.17 & 0.29 $\pm$ 0.01 & 1.44 $\pm$ 0.67 & 0.31 $\pm$ 0.02 \\
N I & 1318.98 & 1318.93 & -10.97 & 1.08 $\pm$ 0.17 & 0.25 $\pm$ 0.01 & 1.6 $\pm$ 0.68 & 0.3 $\pm$ 0.02 \\
N I & 1319.68 & 1319.65 & -6.58 & 0.5 $\pm$ 0.16 & 0.38 $\pm$ 0.02 & 0.72 $\pm$ 0.64 & 0.49 $\pm$ 0.1 \\
S I & 1323.52 & 1323.51 & -4.37 & 0.74 $\pm$ 0.15 & 0.34 $\pm$ 0.02 & 1.28 $\pm$ 0.59 & 0.71 $\pm$ 0.24 \\
C II * & 1323.93 & 1323.9 & -4.37 & 1.52 $\pm$ 0.17 & 0.31 $\pm$ 0.01 & 2.65 $\pm$ 0.67 & 0.37 $\pm$ 0.02 \\
S I & 1326.65 & 1326.61 & -6.54 & 0.61 $\pm$ 0.15 & 0.31 $\pm$ 0.01 & 1.14 $\pm$ 0.6 & 0.3 $\pm$ 0.03 \\
C I & 1328.83 & 1328.83 & 0.0 & 1.35 $\pm$ 0.13 & 0.27 $\pm$ 0.01 & 1.94 $\pm$ 0.54 & 0.32 $\pm$ 0.04 \\
C I & 1329.1 & 1329.08 & -2.18 & 1.89 $\pm$ 0.15 & 0.31 $\pm$ 0.01 & 2.58 $\pm$ 0.6 & 0.39 $\pm$ 0.04 \\
C I & 1329.58 & 1329.57 & 0.0 & 2.17 $\pm$ 0.18 & 0.28 $\pm$ 0.01 & 3.23 $\pm$ 0.73 & 0.33 $\pm$ 0.02 \\
C II * & 1334.53 & 1334.56 & 6.5 & 76.88 $\pm$ 0.39 & 0.24 $\pm$ 0.0 & 131.83 $\pm$ 1.74 & 0.36 $\pm$ 0.01 \\
C II * & 1335.71 & 1335.64 & -15.16 & 149.19 $\pm$ 0.44 & 0.29 $\pm$ 0.0 & 222.51 $\pm$ 1.96 & 0.37 $\pm$ 0.01 \\
C II * & 1351.66 & 1351.64 & -4.28 & 7.39 $\pm$ 0.26 & 0.19 $\pm$ 0.0 & 7.61 $\pm$ 1.01 & 0.21 $\pm$ 0.01 \\
Fe XXI * & 1354.05 & 1354.07 & 2.14 & 7.68 $\pm$ 0.31 & 0.59 $\pm$ 0.01 & 7.83 $\pm$ 1.21 & 0.71 $\pm$ 0.03 \\
C I & 1354.29 & 1354.2 & -21.36 & 7.69 $\pm$ 0.31 & 0.59 $\pm$ 0.01 & 7.87 $\pm$ 1.21 & 0.71 $\pm$ 0.03 \\
C I & 1355.84 & 1355.79 & -8.53 & 7.12 $\pm$ 0.28 & 0.33 $\pm$ 0.01 & 9.31 $\pm$ 1.1 & 0.43 $\pm$ 0.02 \\
C I & 1357.13 & 1357.11 & -4.26 & 1.31 $\pm$ 0.2 & 0.34 $\pm$ 0.01 & 1.85 $\pm$ 0.78 & 0.42 $\pm$ 0.03 \\
C I & 1357.66 & 1357.63 & -6.39 & 0.81 $\pm$ 0.18 & 0.36 $\pm$ 0.02 & 1.26 $\pm$ 0.7 & 0.43 $\pm$ 0.05 \\
C I & 1359.28 & 1359.27 & -2.13 & 1.66 $\pm$ 0.22 & 0.41 $\pm$ 0.02 & 2.29 $\pm$ 0.87 & 0.46 $\pm$ 0.03 \\
\enddata
\tablecomments{Flux measurements are given in $10^{-15}$ erg/s/cm$^{-2}$. Ions marked with * are optically thick lines used when fitting the DEM models following the methods of \cite{duvvuri21}.}
\end{deluxetable}\label{tab:linelist}

\software{
 numpy \citep{numpy},
 matplotlib \citep{matplotlib},
 scipy \citep{2020SciPy}
 astropy \citep{astropy:2013, astropy18},
 \costools\footnote{https:\/\/github.com\/spacetelescope\/costools},
 \calcos\footnote{https:\/\/github.com\/spacetelescope\/calcos},
 \texttt{lmfit} \citep{lmfit},
 \texttt{calcos}/\texttt{costools} \citep{soderblom21},
 PlasmaPy \citep{plasmapy},
 ChiantiPy\footnote{\url{https://github.com/chianti-atomic/ChiantiPy}}
 }

\facility{Hubble Space Telescope,
   Far Ultraviolet Spectroscopic Explorer \citep{moos00, sahnow00},
   XMM-Newton,
   International Ultraviolet Explorer \citep{Boggess78},
   HARPS-N \citep{mayor03},
   }

\bibliography{exopapers}

\end{document}